\documentclass[aps,floats,twocolumn,prl,showpacs,10pt]{revtex4-1}
\usepackage{graphicx}
\usepackage{epstopdf}
\usepackage[colorlinks=true,urlcolor=blue,citecolor=blue,linkcolor=blue,breaklinks=true]{hyperref}
\usepackage{color}
\usepackage{colortbl}
\usepackage{xcolor}
\usepackage{setspace}
\usepackage[utf8]{inputenc}
\usepackage{amsmath}
\usepackage{placeins}
\usepackage{amssymb}
\usepackage{nicefrac}

\begin{document}

\title{Fingerprints of the local moment formation and its Kondo screening \\ in the generalized susceptibilities of many-electron problems }

\author{P. Chalupa$^a$}
\author{T. Sch\"afer$^{b,c}$}
\author{M. Reitner$^{a}$}
\author{D. Springer$^{a}$}
\author{S. Andergassen$^{d}$}
\author{A. Toschi$^a$}

\affiliation{$^a$Institute of Solid State Physics, TU Wien, A-1040 Vienna, Austria}
\affiliation{$^b$Coll{\`e}ge de France, 11 place Marcelin Berthelot, 75005 Paris, France}
\affiliation{$^c$CPHT, CNRS, {\'E}cole Polytechnique, Institut Polytechnique de Paris, Route de Saclay, 91128 Palaiseau, France}
\affiliation{$^d$Institut für Theoretische Physik and Center for Quantum Science, Universität T{\"u}bingen, Auf der Morgenstelle 14, 72076 T{\"u}bingen, Germany}

\date{ \today }

\begin{abstract} 
We identify the precise hallmarks of the local magnetic moment formation and its Kondo screening in the frequency structure of the generalized charge susceptibility. The sharpness of our identification even pinpoints an alternative criterion to determine the Kondo temperature of strongly correlated systems on the two-particle level, which only requires calculations at the {\sl lowest} Matsubara frequency.
We showcase its strength by applying it to the single impurity and the periodic Anderson model as well as to the Hubbard model. 
Our results represent a significant progress for the general understanding of quantum field theory at the two-particle level and allow for tracing the limits of the physics captured by perturbative approaches in correlated regimes.
\end{abstract}

\maketitle

{\sl Introduction.}
The goal of any successful theory is to extract essential 
features of the phenomena of interest from the complexity of the physical world, 
neglecting all superfluous pieces of information. 
This objective is particularly crucial for the cutting-edge quantum field theory (QFT) approaches
designed to describe many-electron systems in the presence of
strong correlations.

Presently, one can rely on a solid textbook interpretation\cite{Abrikosov1975,Mahan2000} of the QFT formalism describing the single-particle (1P) processes, 
measurable e.g.~by (angular resolved) direct and inverse photoemission\cite{PhotoRMP2003} 
or scanning tunneling microscopy\cite{STMRMP1987,STMRMP2007}. 
Crucial information about the metallic or insulating nature of a given many-electron problem, as well as quantitative information about the electronic mass renormalization $Z$ and quasiparticle lifetime $\tau$ is encoded in the momentum/energy dependence of the electronic self-energy $\Sigma$.
If the temperature $T$ is low enough, even a quick glance at the low-energy behavior of $\Sigma$, either in real or in Matsubara frequencies, yields a qualitatively reliable estimate of the most important physical properties. 

The situation is clearly different on the two-particle (2P) level, which can be experimentally accessed by e.g. inelastic neutron scattering\cite{INS,INS2}. 
Due to the complex physical mechanisms at play, the related textbook knowledge is mostly limited to general 
definitions\cite{Abrikosov1975,Mahan2000}. For this reason, corresponding analytical/numerical calculations are often performed with significant approximations or with a black-box treatment of the 2P processes.
However, the last decade has seen a rapid development of methods at the forefront of the many-electron theory\cite{Maier2005,Metzner2012,Rohringer2018}, for 
which generalized 2P correlation functions are the key ingredient. This is reflected in an increasing 
effort to develop 
the corresponding formal aspects and algorithmic 
procedures\cite{Kunes2011,Rohringer2012, Metzner2012, Hafermann2014, Gunnarsson2015, Gunnarsson2016, Wentzell2016, Kaufmann2017,Kugler2018, Tagliavini2018,Stepanov2018a,vanLoon2018,Rohringer2018,Maier2005,Kugler2018b,  Stepanov2018b, Krien2019SBE,Stepanov2019,Krien2019SBEb,Hille2020,vanLoon2020,Reitner2020}.
At the same time, the rather poor physical understanding of the 2P processes remains largely behind 
the requirements of the most advanced QFT methods.
Interesting progress has been recently reported\cite{Krien2019,Kotliar2020} on the relation of 1P Fermi-liquid parameters to 2P scattering functions.
Ideally, however, one would like to be 
able to interpret 
the physics encoded at the 2P level
with a similar degree of confidence as  
for the 1P processes.


In our paper, we make a significant step forward in this direction: We identify the 
fingerprints of two major 
hallmarks of strong correlations  
in the generalized charge 
susceptibility.
In particular, we pinpoint the frequency structures encoding 
the formation of local magnetic moments as well as of their Kondo screening.
In this perspective, we also show how the Kondo temperature $T_{\rm K}$ corresponds  
to a specific property of the generalized charge susceptibility, allowing for an alternative, simple 
path of extracting its value directly from the lowest Matsubara frequency data.
\begin{figure*}[t]
\centering
{{\resizebox{17.5cm}{!}{\includegraphics {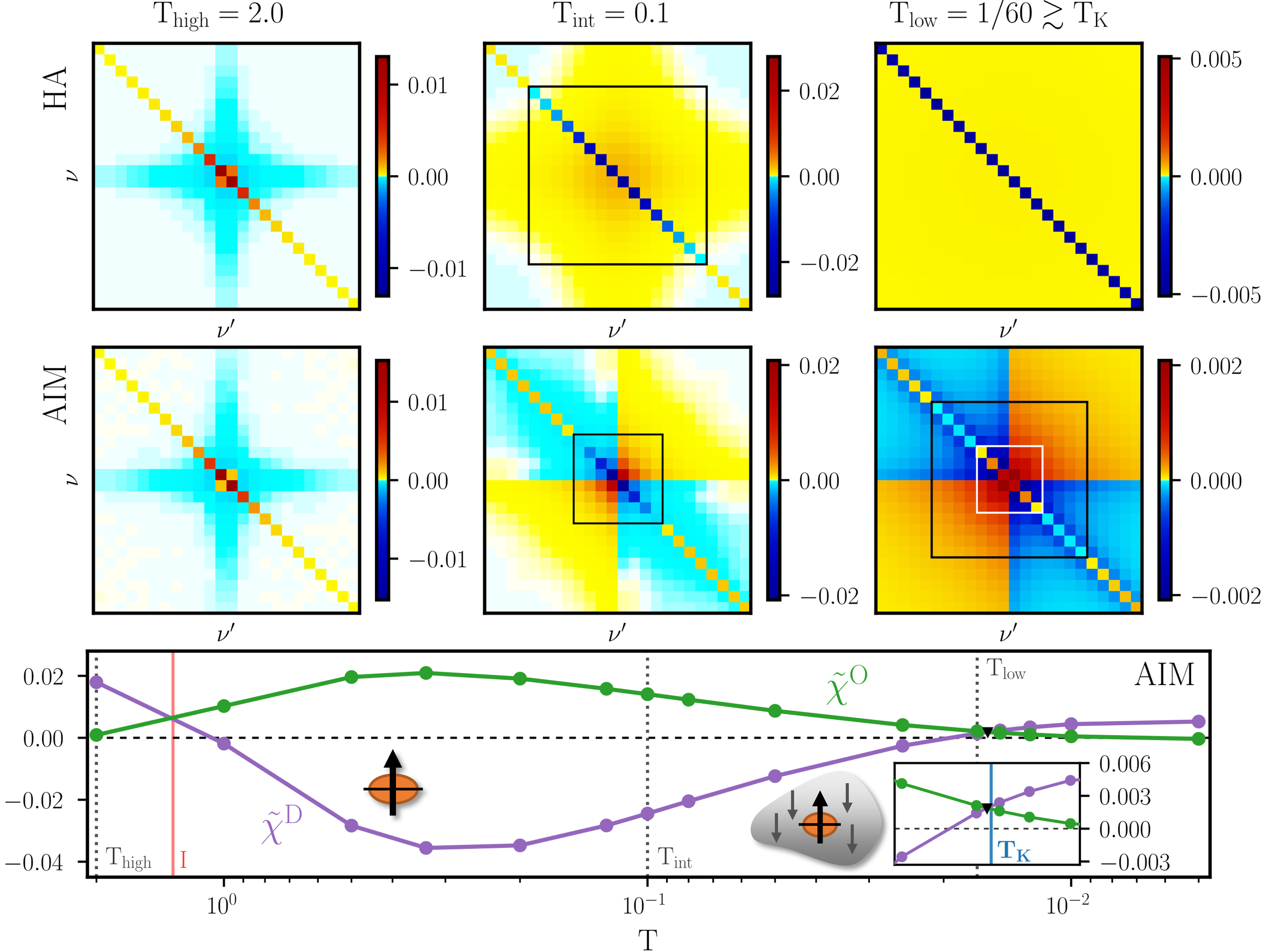}}} 
\caption{{\small Comparison of the Matsubara frequency structure of $T^2\tilde{\chi}^{\nu\nu'}(\Omega\!=\!0)$ for the HA (top row) and the AIM (center row) for 
$U\!=\!5.75$\cite{UNote} and different temperatures. The maximal Matsubara index is kept fixed for all temperatures (the labels are hidden to ensure better readability). Black and white squares mark the main frequency structures, as described in the text. 
Lower panel: Temperature evolution of the lowest Matsubara frequency elements of $T^2\tilde{\chi}^{\nu\nu'}(\Omega\!=\!0)$: 
$\tilde{\chi}^{\rm D}\!=\!T^2\tilde{\chi}^{\,\pi T, \, \pi T}$ (violet) and $\tilde{\chi}^{\rm O}\!=\!T^2\tilde{\chi}^{\,\pi T, -\pi T}$ (green). They cross at $T_{\text{high}}$ at the divergence of $\Gamma$ (red (I)), and at low-temperatures at 
$T\simeq T_{\rm K}$ (black triangle), see also the inset showing a zoom around $T_{\rm K}$ (vertical blue line). 
The arrows with/without the surrounding cloud sketch the local moment/the Kondo screened regime.  
}} 
\label{fig:1}  
}
\end{figure*}

We recall that the Kondo problem\cite{Hewson1993} 
provides a paradigm for a variety of physical effects\cite{FISK33,SC1,SC2,coleman2006heavy,Andergassen2010} involving strong electronic correlations.  
Local moment formation and Kondo screening are also a crucial ingredient of the physics described by the dynamical mean-field theory (DMFT)\cite{Georges1996} through the solution of a 
self-consistently determined auxiliary Anderson impurity model (AIM).


Learning how to extract important 
physical information from the generalized susceptibility represents a substantial improvement for the understanding of quantum many-electron physics at the 2P level.
Further, having this information at hand also enables us to draw 
conclusions on 
two relevant theoretical questions: (i) The relation of the recently reported multifaceted manifestations\cite{Gunnarsson2017} of the breakdown of perturbation theory, such as the divergences of the irreducible vertex functions\cite{Schaefer2013, Janis2014,Ribic2016, Schaefer2016c,Gunnarsson2016,Vucicevic2018, Chalupa2018,Thunstroem2018,Springer2019} and the crossing of multiple solutions\cite{Kozik2015, Stan2015, Schaefer2016c, Gunnarsson2017, Tarantino2018, Thunstroem2018, Vucicevic2018} of the Luttinger-Ward functional, with the local moment physics and its Kondo screening; (ii) the built-in limit of advanced perturbative 
approaches to describe these fundamental physical effects.

\begin{figure*}[ht!]
\centering
{{\resizebox{17.8cm}{!}{\includegraphics {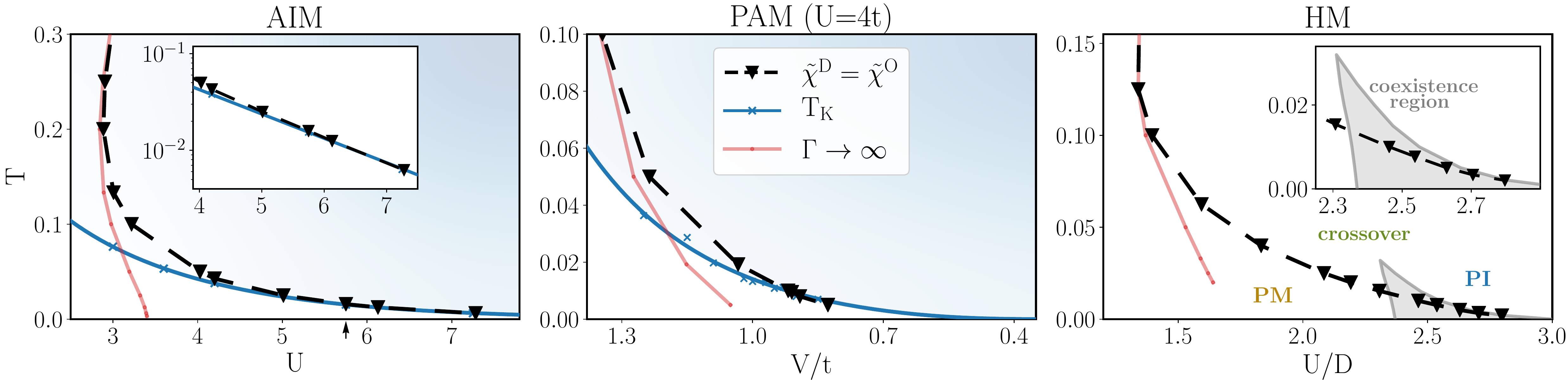}}} 
\caption{{\small Phase diagram of the AIM (left), the PAM (DMFT) (middle), and the HM (DMFT) (right) as a function of the 
interaction $U$ (hybridization $V$ for the PAM, $U$ fixed) and the temperature $T$ showing the line where $\tilde{\chi}^{\rm D}\!=\!\tilde{\chi}^{\rm O}$ holds (black triangles dashed), 
i.e. the singularity of the $2\!\times\!2$ submatrix of $\tilde{\chi}^{\nu \nu'}$. The left and central panels show the agreement at low temperatures between $T_{\rm K}$ (blue solid line) and 
the condition $\tilde{\chi}^{\rm D}\!=\!\tilde{\chi}^{\rm O}$, clearly evident also in logarithmic scale (left inset). The local moment regime is represented by a bluish shadowed area in both panels. 
The red lines denote the (first) divergence of the irreducible vertex $\Gamma$. For the HM on the Bethe lattice the paramagnetic metallic (PM)/insulating (PI) phases are indicated together with their crossover. 
The coexistence region is shown in gray. The arrow on the abscissa (left) marks the interaction value used in Fig.~\ref{fig:1} and \ref{fig:3}. }}
\label{fig:2}  
}
\end{figure*}

{\sl How to read two-particle quantities.}  
We start from the definition of the generalized local susceptibility\cite{Rohringer2012,Rohringer2018, Suppl}
\begin{equation}
\tilde{\chi}_{\sigma\sigma'}^{\, \nu\nu'}(\Omega) \! = \! G^{(2)}_{\sigma\sigma'}(\nu, \nu',\Omega) \! - \! T^{-1}G(\nu)G(\nu')\delta_{\Omega
0}\delta_{\sigma\sigma'} 
\label{eq:chigen}    
\end{equation}
in terms of the 2P ($G^{(2)}$) and 1P ($G$) Green's  
functions, where  $\nu,\nu'$ and $\Omega$ are fermionic and bosonic Matsubara frequencies, respectively, and $\sigma,\sigma'\!=\!\{\uparrow,\downarrow\}$  spin indices.
 As we show in the following for repulsive interactions, 
 the generalized {\sl charge} susceptibility $\tilde{\chi}^{\,\nu \nu'}(\Omega)\!=\!\tilde{\chi}_{\uparrow \uparrow}^{\, \nu\nu'}(\Omega) + \tilde{\chi}_{\uparrow\downarrow}^{\, \nu\nu'}(\Omega)$ 
 allows for the best readability of the underlying physics at the 2P level. Furthermore, the physical response of this sector captures 
 the fundamental properties of any interacting electron system. 
 We recall that the physical response function ($\chi$) is obtained from the generalized susceptibility $\tilde{\chi}^{\nu \nu'}(\Omega)$ by summing over the fermionic Matsubara frequencies $\nu$, $\nu'$\cite{Suppl}. 
 The static charge response $\chi(\Omega\! = \! 0)$ reads
 \begin{equation}
 \chi = T^2 \sum\limits_{\nu\nu'}\tilde{\chi}^{\, \nu\nu'} = T^2 \sum\limits_{\nu\nu'}{(\tilde{\chi}_{\uparrow\uparrow}^{\, \nu\nu'}+\tilde{\chi}_{\uparrow\downarrow}^{\, \nu\nu'})}.
 \label{eq:chiphys}
 \end{equation}
We start by analyzing the arguably simple case of an isolated atom with a repulsive interaction $U$ (Hubbard atom, HA), where analytic expressions are also available\cite{Rohringer2012, Thunstroem2018}.
This represents the purest realization of local moment physics, which hence provides an ideal baseline for the interpretation of the more interesting cases discussed below.  
In Fig.~\ref{fig:1} (upper panels), we show an 
intensity plot of $\tilde{\chi}^{\nu\nu'}$ (normalized by $T^2$) for 
$U\!=\!5.75$\cite{UNote}, half filling (where $\tilde{\chi}^{\nu \nu'}$ is real\cite{Rohringer2012,Thunstroem2018}) and different temperatures. 
At 
high temperature 
($T_{\text{high}}\!=\!2$, left panel), the overall frequency structure consists of a large positive-valued diagonal (yellow/red) and a weak negative cross structure (blue). This corresponds to 
a typical {\sl perturbative} behavior\cite{Rohringer2012,Wentzell2016}, 
dominated by the diagonal bubble term
$\tilde{\chi}_0^{\, \nu\nu'} \!=\! \tilde{\chi}_{0, \uparrow \uparrow}^{\, \nu\nu'}  \!= \! - \delta_{\nu \nu'} G(\nu)^2/T $: 
Correlation effects are washed out for $T  \! \gtrsim \! U$, 
consistent with the 
feasibility of
high-$T$ expansions.

The situation changes radically when reducing $T$: in the intermediate ($T_{\rm int}\!=\!0.1$) and low ($T_{\rm low}\!=\!1/60\!\approx\!0.017$) temperature regime (central and right panel), one observes a strong {\sl damping} of all diagonal elements of $\tilde{\chi}^{\nu\nu'}$. The effect is more pronounced at low frequencies, as the sign of $\tilde{\chi}^{\nu=\nu'}$ becomes even {\sl negative} (bluish colors) for $|\nu|\!\lesssim\!U$\cite{Thunstroem2018} (black square). 
This major feature is accompanied by the appearance of small {\sl positive} off-diagonal elements (yellow). 
The net effect is a 
suppression of the physical susceptibility $\chi$, see Eq.~(\ref{eq:chiphys}), 
which occurs when the thermal energy is no longer large enough ($T\!\sim\!\nu\!<\!U$) to counter the formation of a local moment driven by $U$, eventually 
yielding an exponentially small $\chi \sim \! e^{-U/2T}$ for $T\rightarrow0$. 
Altogether, the low-$T$ HA results illustrate how the onset of a pure local moment is encoded in the 
charge sector: a progressive emergence of a nonperturbative sign structure in $\tilde{\chi}^{\nu \nu'}$, which is the opposite image of the perturbative one (left panel). This also induces
several negative eigenvalues of $\tilde{\chi}^{\nu\nu'}$, responsible for the breakdown of perturbative expansions\cite{Gunnarsson2017}.


Let us now examine how this picture changes when the HA system is connected to an electronic bath (here: with a flat DOS of bandwidth $W\!=\!20$ and 
hybridization $V\!=\!2\!<\!U\!=\!5.75$\cite{UNote}), corresponding to the well-known
Anderson impurity model (AIM). 
By comparing the results of $T^2\tilde{\chi}^{\nu\nu'}$ (central-row panels of Fig.~\ref{fig:1}, computed
with w2dynamics\cite{Suppl}) 
to those of the HA, we observe almost no difference at $T_{\rm high}$. This is not surprising as thermal fluctuations prevail over 
both correlation ($U$) and hybridization ($V$) effects in this case.
Upon lowering $T$ to $T_{\rm int}$, 
we enter the {\sl local moment} regime of the AIM. This is reflected in 
a qualitatively similar evolution as seen in the HA:
a progressive suppression of the diagonal entries of $\tilde{\chi}^{\nu\nu'}$, turning negative in the low-energy sector (black square), 
accompanied by positive, yet smaller, off-diagonal contributions, with an overall freezing effect on the local density fluctuations (see Eq.(\ref{eq:chiphys}) and \cite{Suppl}).
This is how the formation of a local moment affects the charge sector, thus representing its {\sl fingerprint}. 
However, due to the screening effects of the bath its features get weakened, explaining the quantitative differences to the HA (e.g., the reduced size 
of the black square). 

The most interesting situation is encountered when reducing $T$ further down to $T_{\rm low}\gtrsim T_{\rm K}$ (right panel), where the Kondo screening induces 
{\sl qualitative} differences w.r.t.~the HA. We observe that the low-frequency diagonal elements of $\tilde{\chi}^{\nu \nu'}$ (white square) are flipped back to positive, as in the perturbative regime.
This trend is driven by 
the low-energy correlations between electrons with antiparallel spins ($\tilde{\chi}_{\uparrow\downarrow}^{\, \nu=\nu'}$)\cite{Suppl}. The weakening of their negative contribution increases the physical 
charge susceptibility $\chi$ (see Eq.~(\ref{eq:chiphys}) and \cite{Suppl}) and simultaneously mitigates 
the magnetic response.
However, in the intermediate frequency regime, the diagonal elements of $\tilde{\chi}^{\nu \nu'}$ are still negative, reflecting the underlying presence of a (partially screened) local moment. 
The fingerprint of the Kondo regime is, thus, the {\sl onion}-like frequency structure of $\tilde{\chi}^{\nu \nu'}$, which is clearly recognizable in the rightmost central panel of Fig.~\ref{fig:1}: (i) a high-frequency perturbative asymptotic, (ii) a local moment driven structure (with suppressed diagonal) at intermediate frequencies, (iii) an inner core (with a similar sign structure as (i)) induced by the Kondo screening. 
A quick glance at the sign structure of $\tilde{\chi}^{\nu\nu'}$ therefore allows for an immediate understanding of the underlying physics. 
This nicely illustrates the balanced competition in the charge sector between the freezing effects of the local moment and the defreezing effects of its low-energy screening, which characterizes the Kondo regime.

Note, that the onion-structure is also found for other values of $U$, as well as in other models\cite{Suppl}, discussed below. 

{\sl How to extract the Kondo temperature.} 
The behavior described above is also reflected in the temperature evolution of the lowest frequency entries of $\tilde{\chi}^{\nu\nu'}$: 
the diagonal $\tilde{\chi}^{\rm D}\!=\!T^2\tilde{\chi}^{\, \pi T, \, \pi T}$ and the off-diagonal 
$\tilde{\chi}^{\rm O}\!=\!T^2\tilde{\chi}^{\, \pi T, -\pi T}$, shown in the lowest panel of Fig.~\ref{fig:1}. 
We can readily trace the sign changes marking the three regimes discussed above, associating 
the (negative) minimum of $\tilde{\chi}^{\rm D}$ with the temperature at which the strongest local moment effects are observed. 
The screening induced enhancement of $\tilde{\chi}^{\rm D}$ at lower temperatures has {\sl remarkable consequences}: We find that crossing the Kondo temperature, 
as defined in a standard way from the behavior of the static magnetic response of the system\cite{Suppl}
($T_{\rm K}\!=\!1/65\!\approx\!0.015$ at $U\!=\!5.75$ for the AIM),
matches with high accuracy the equality of $\tilde{\chi}^{\rm D}$ and $\tilde{\chi}^{\rm O}$ observed at low-$T$ (s.~inset of Fig.~\ref{fig:1}, marked by black triangle).
We emphasize that this criterion holds more generally. 
As shown in the phase diagram of the AIM in Fig.~\ref{fig:2} (left panel), 
the condition $\tilde{\chi}^{\rm D}\!=\!\tilde{\chi}^{\rm O}$ (black triangles) perfectly traces $T_{\rm K}$ (blue line)\footnote{For different interaction values $T_{\rm K}$ is 
extracted\cite{Suppl} (blue crosses), and then fitted using an analytical expression for $T_{\rm K}$ in the wide-band limit\cite{Hewson1993} 
($A\sqrt{U\Delta}\exp{(-B\frac{U}{\Delta} + C\frac{\Delta}{U})}$).} in the {\sl entire} local moment regime $T, V < U$ (see also the logarithmic inset), i.e. where the definition of a Kondo scale is actually meaningful. Note that this is not the case for other criteria one could naturally think of, such as $\tilde{\chi}^{\rm D}\!=\!-\tilde{\chi}^{\rm O}$ or $\tilde{\chi}^{\rm D}\!=\!0$\cite{Suppl}.

Moreover, 
our simple 2P definition of $T_{\rm K}$ holds also beyond the single impurity problem. 
In Fig.~\ref{fig:2}, we show DMFT calculations for the periodic Anderson model on a square lattice with nearest-neighboring hopping $t$ (PAM, central) and for a 
Hubbard model on a Bethe lattice with unitary half-bandwidth $D$ (HM, right)\cite{Suppl}. 

In particular, we observe that for the PAM, the {\sl same matching} of the condition 
$\tilde{\chi}^{\rm D}\!=\!\tilde{\chi}^{\rm O}$ (black triangles) and $T_{\rm K}$~\cite{Suppl,Schaefer2019,ladderDGA} (blue line) is found in the local moment regime (i.e., when $V < t$, blue-shadowed area). 

In the HM, the Kondo temperature characterizing the auxiliary AIM associated with the self-consistent DMFT solution, depends on the temperature itself: $T_{\rm K}^{\rm \, HM}(T)$. 
Hence, $\tilde{\chi}^{\rm D}\!=\!\tilde{\chi}^{\rm O}$ (black triangles) indicates 
that the temperature equals the effective Kondo temperature, i.e.~$T_{\rm K}^{\rm \, HM}(T)\!=\!T$. Physically, it is natural to relate this condition to 
the onset of low-energy electronic coherence: For all temperatures below the $\tilde{\chi}^{\rm D}\!=\!\tilde{\chi}^{\rm O}$ condition, 
a conventional Fermi-liquid behavior of the physical response can be expected 
(e.g.: $\rho(T) \propto T^2, c_V(T) \propto T$,  etc.\cite{Abrikosov1975}). 
This would also be 
consistent with the $\tilde{\chi}^{\rm D}\!=\!\tilde{\chi}^{\rm O}$ condition 
approaching the Mott Hubbard metal-insulator transition (MIT) at $U_{\rm MIT}(T\!=\! 0)\!=\!U_{c2}$ in the low-$T$ limit 
(see also recent DMFT studies of the physics in the proximity of the MIT\cite{Terletska2011,Vucicevic2013}). 
%
%

The equality of the elements of the innermost 
$2\times 2$ submatrix of $\tilde{\chi}^{\nu \nu'}$ represents therefore a very simple, clear-cut criterion for determining $T_{\rm K}$ at the 2P level.

{\sl A non-perturbative Fermi liquid.}
Beyond its physical relevance, our improved 2P understanding sheds light onto the nontrivial relation with the breakdown of perturbation theory\cite{Gunnarsson2017}. 
At high $T$, where $\nu_0 \!=\! \pi T \gtrsim V, U, t$, the $2\times 2$ 
submatrix encodes all 
relevant energy scales, the rest being nonsingular high-frequency asymptotics. 
In this case $\tilde{\chi}^{\rm D}\!=\!\tilde{\chi}^{\rm O}$ corresponds to a singular eigenvalue of the {\sl entire}  $\tilde{\chi}^{\nu \nu'}$
and hence to 
a divergence of the irreducible vertex function $\Gamma^{\nu \nu'}\!=\! [\tilde{\chi}^{\nu\nu'}]^{-1}\!-\![\tilde{\chi}_0^{\nu\nu'}]^{-1}$, specifically 
to the first (I) one encountered when reducing the temperature (red line in Figs.~\ref{fig:1} and \ref{fig:2})\cite{Schaefer2013,Schaefer2016c,Chalupa2018,Thunstroem2018,Springer2019}.
\begin{figure}[t!]
\centering
{{\resizebox{8.4cm}{!}{\includegraphics {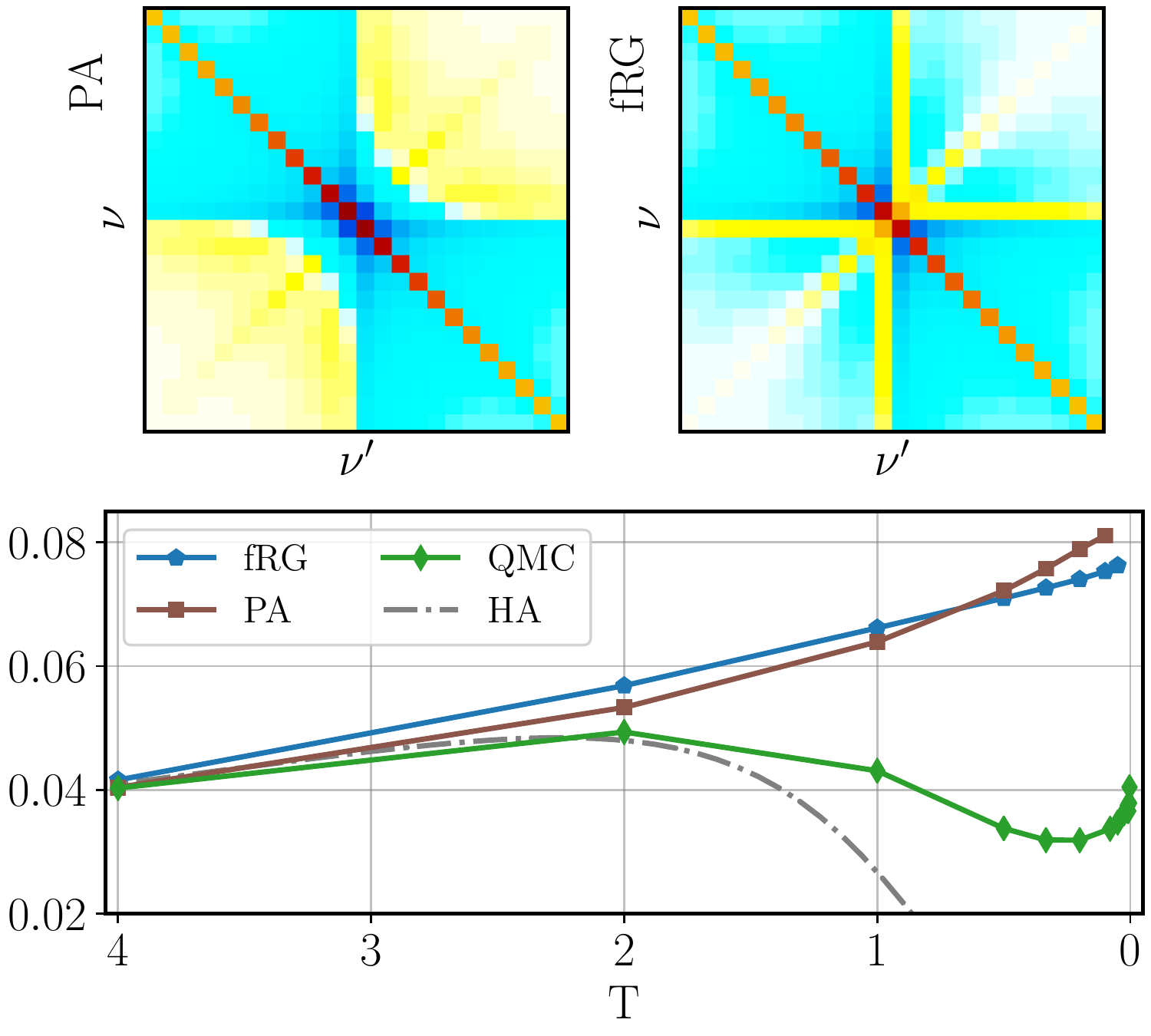}}}
\caption{{\small Generalized charge susceptibility ($T^2\tilde{\chi}^{\nu\nu'}$) for the AIM, 
as obtained by means of PA and fRG for $U\!=\!5.75$ and $T\!=\!T_{\rm int}$. The same color scale as in Fig.~\ref{fig:1} 
(AIM, $T_{\rm int}$) is used. 
Lower panel: static physical charge susceptibility $\chi$ computed with different approaches as a function of $T$.
}} 
\label{fig:3}
}
\end{figure}
For intermediate temperatures, the $2\!\times\!2$ submatrix is controlled by the local moment, leading to a strongly negative $\tilde{\chi}^{\rm D}$ and negative eigenvalues of the submatrix (as in the HA case). 
At $T_{\rm K}$ the eigenvalue flips sign and one finds again $\tilde{\chi}^{\rm D} > \tilde{\chi}^{\rm O}$ for $T \lesssim T_{\rm K}$, as in the perturbative regime (s. Fig.~\ref{fig:1}, lowest panel).
Here, however, because of the onion-like structure of $\tilde{\chi}^{\nu \nu'}$, the positive-definiteness (and thus the invertibility) is guaranteed only for an inner submatrix 
describing the Fermi liquid regime, but not for the full 
$\tilde{\chi}^{\nu \nu'}$. This explains why divergences of irreducible vertex functions can occur also at 
low temperatures\cite{Chalupa2018} even in the presence of a Fermi liquid ground state.  Indeed, such vertex divergences mark the distinction between a Fermi liquid in the weak- and in the strong-coupling regime. 

{\sl Limitations of perturbative approaches.}  
The direct link between the 2P fingerprints of local moments and vertex divergences, sets precise physical limitations for perturbative methods, where -per construction- $\Gamma$ is finite\footnote{With the only exception of second-order phase transitions to long-range ordered phases, not relevant here.}.
Hence, the impact of the characteristic physics emerging from the magnetic sector onto the charge channel, cannot be described 
by perturbative methods.
We substantiate this statement by considering two advanced perturbative schemes, the functional renormalization group (fRG)\cite{Metzner2012,Suppl}
and the parquet approximation (PA)\cite{Bickersbook2004,Janis2007,Janis2008,Yang2009,Tam2013,Valli2015,Wentzell2016,Li2016,Janis2019,Janis2019JPS,Kauch2020,Janis2020,Suppl}.
The results obtained for the AIM with $U\!=\!5.75$\cite{UNote} and 
$T \! =\! T_{\rm int}$ are shown in Fig.~\ref{fig:3}. $\tilde{\chi}^{\nu \nu'}$ computed by the fRG and PA (upper panels) appear qualitatively different from the (numerically) exact one of Fig.~\ref{fig:1} (AIM, central): The diagonal elements are all {\sl positive} and substantially larger than the off-diagonal ones. 
This ensures the positive-definiteness of the entire $\tilde{\chi}^{\nu \nu'}$, preventing the suppression effects of the charge response, which characterize the local moment regime. 
This drawback qualitatively affects the physical description. 
%
In particular, the temperature dependence of the numerically exact physical charge susceptibility $\chi$ (Fig.~\ref{fig:3} lower panel) exhibits a {\sl clear minimum} 
for intermediate $T_{\rm high} \!>\! T \!>\! T_{\rm K}$.  
This emerges from the competition between the suppression induced by the local moment (see the extreme HA case) and the low-energy screening.   
Both features are {\sl not} captured by the fRG (blue pentagons) and PA (brown squares), which 
display a monotonous behavior as $T$ is decreased, in the framework of a mere thermal quenching. 
At the same time, the perturbative approaches are able to capture the qualitative correct behavior of the magnetic response, 
reflecting the absence of divergences of $\Gamma$ in this sector\cite{Suppl}.


{\sl Conclusions.} We have shown how fundamental physical properties of correlated systems, i.e.~the local moment formation and its Kondo screening, can be directly read from
the Matsubara frequency structure of the generalized {\sl charge} susceptibility $\tilde{\chi}^{\nu \nu'}$. 
In particular, the competition between localization effects at higher energies and metallic screening at lower energies is encoded in a clearly recognizable {\sl ``onion-like"} fingerprint of $\tilde{\chi}^{\nu \nu'}$, 
emerging in the Kondo regime.
The thorough inspection of the latter even discloses an alternative route to extract $T_{\rm K}$ from the charge-sector.
Our improved understanding of the 2P-processes sets also clear-cut limits to the physics accessible to perturbative approaches. 

As a future perspective, it will be worth to overcome the on-site/single-orbital framework of our study.
We expect that the role of the local moments will be played by short-range\cite{Maier2005, Sordi2010,Sordi2012,Gunnarsson2015}/Hund's-driven\cite{Haule2009,DeMedici2011,Chubukov2015} magnetic fluctuations. Their nonperturbative images could reverberate, analogously as presented here, onto the charge/pairing response of the system. The identification of the corresponding fingerprints may open new pathways toward a microscopic understanding of unconventional superconductivity in the nonperturbative regime.\\

\begin{acknowledgments}
{\sl Acknowledgments} - We thank M. Capone, S. Ciuchi, J. von Delft, K. Held, C. Hille, F. Krien, F.B. Kugler, E. van Loon, C. Schattauer, and G. Sangiovanni for insightful discussions. 
The authors also want to thank the CCQ of the Flatiron Institute (Simons Foundation) for the great hospitality. The present work was supported by the Austrian Science Fund (FWF) through the Project I 2794-N35 and Erwin-Schr\"odinger Fellowship J 4266 - ``{\sl Superconductivity in the vicinity of Mott insulators}'' (SuMo, T.S.), the Deutsche Forschungsgemeinschaft (DFG) through Project No. AN 815/6-1, as well as the European Research Council for the European Union Seventh Framework Program (FP7/2007-2013) with ERC Grant No. 319286 (QMAC, T.S.).
\end{acknowledgments}

\bibliography{Fingerprints}

\begin{thebibliography}{107}
\bibitem [{UNo()}]{UNote}%
  \BibitemOpen
  \href@noop {} {}\bibinfo {note} {The value of $U\!=\!5.75$ has been chosen
  such that the local moment is already well defined and, at the same time,
  calculations are still feasible within PA and fRG, allowing for direct
  comparisons. Similar results are obtained for other $U$'s, see \cite{Suppl}.
  For the AIM the interaction is often compared to $\Delta\!=\!V^2\pi\rho_0$.
  In our case $\rho_0=\frac{1}{W}$ and $\Delta\!=\!\pi/5$, hence
  $U/\Delta\!\approx\!9.15$.}\BibitemShut {Stop}%
\end{thebibliography}

\clearpage

\section*{SUPPLEMENTAL MATERIAL}


In this supplemental material we provide specific definitions as well as
technical details on the numerical methods (impurity solver, DMFT, determination of the Kondo temperature $T_{\rm K}$) 
applied for the calculations presented in the main text. 
We also append additional details on the frequency structure of the generalized susceptibility. In the last part we briefly recall the essential features of the perturbative schemes used for comparison and 
analyze the limitations of these approaches in more detail.


\section{Formalism and numerical methods}

\subsection{General definitions}

The explicit definition of the local
generalized susceptibility (see Eq.~(1) of the main text), consistently with the notation of Ref.~[\onlinecite{Rohringer2012}], 
reads
\begin{eqnarray}
\tilde{\chi}^{\,\nu\nu'\Omega}_{\sigma\sigma'} & = & \int d\tau_1 d\tau_2 d\tau_3 \,
e^{-i\nu\tau_1} e^{i(\nu+\Omega)\tau_2} e^{- i(\nu'+\Omega)\tau_3}   \nonumber\\ \nonumber
    & \times & \left[ \langle T_\tau c^\dagger_\sigma(\tau_1) c_\sigma(\tau_2)
    c^\dagger_{\sigma'}(\tau_3) c_{\sigma'}(0)\rangle  \right. \\ 
    &    -  & \left.  \langle T_\tau c^\dagger_\sigma(\tau_1)
    c_\sigma(\tau_2) \rangle
     \langle T_\tau   c^\dagger_{\sigma'}(\tau_3) c_{\sigma'}(0)\rangle \right],
\label{eq:chi}
\end{eqnarray}
where $T_\tau$ is the (imaginary) time ordering operator, 
$\nu,\nu'$ and $\Omega$ denote the two fermionic and the
bosonic Matsubara frequencies, respectively. $c_{\sigma}^{(\dagger)}(\tau)$ annihilates (creates) a particle at imaginary time $\tau$ with spin $\sigma  \in \{\uparrow, \downarrow\}$.
In the main text we consider the static charge channel $\tilde{\chi}^{\,\nu\nu'}\!:=\!\tilde{\chi}^{\,\nu\nu'\Omega\!=\!0}_{\uparrow\uparrow}
\!+\!\tilde{\chi}^{\,\nu\nu'\Omega\!=\!0}_{\uparrow\downarrow}$, in this supplemental material we also report data on the physical magnetic one
$\chi_{m}(\Omega=0)\!:=\!  \int_0^{\beta}\langle S_z(\tau)S_z \rangle \!=
\!\nicefrac{1}{2}\sum_{\nu\nu'}(\tilde{\chi}^{\,\nu\nu'\Omega\!=\!0}_{\uparrow\uparrow}
\!-\!\tilde{\chi}^{\,\nu\nu'\Omega\!=\!0}_{\uparrow\downarrow})$, 
where $S_z=\nicefrac{1}{2}(n_{\uparrow}-n_{\downarrow})$ and $n_{\sigma} = c_{\sigma}^{\dagger}c_{\sigma}$. 
We recall that, for the particle-hole symmetric half-filled case we 
analyze throughout this work, $\tilde{\chi}^{\nu\nu'}$ is real-valued only\cite{Rohringer2012}.

\subsection{Details on the numerical methods}

The calculations for the single impurity Anderson model (AIM) (Hamiltonian see Ref.~\cite{Chalupa2018}; box-shaped density of states 
of the itinerant electrons with a bandwidth of $W\!=\!20$, an energy- and momentum-independent 
hybridization parameter $V\!=\!2$) as well as the DMFT calculations of the Hubbard model (HM) (Bethe lattice with unitary half-bandwidth $D$) were performed using the w2dynamics\cite{w2dynamics} package. This provides an implementation 
of a continuous-time quantum Monte Carlo (CT-QMC)\cite{Gull2011a} solver in the hybridization expansion, which we applied for obtaining one- and two-particle quantities.
In particular, the computations were performed on the Vienna Scientific Cluster (VSC) where we used about $10.000-15.000$ CPU hours for each two-particle measurement, depending on the temperature. The results 
for the physical susceptibilities were obtained from direct independent measurements using w2dynamics\cite{w2dynamics}.

The DMFT calculations of the periodic Anderson model (PAM) (Hamiltonian see Ref.~\cite{Schaefer2019})
were performed using an exact-diagonalization solver for $U=4t$, on the half-filled unfrustrated square lattice.
As described in detail in the Supplemental Material of \cite{Schaefer2019}, in order to obtain the generalized susceptibility on the impurity, a typical number of $N_{\nu}\!=\!120$ positive Matsubara frequencies has been used. We utilized exact diagonalization with four bath sites and one impurity site $N_{s}\!=\!5$ as a solver for the auxiliary Anderson impurity model. The results have also been carefully crosschecked with those obtained by CT-QMC \cite{w2dynamics}.

\section{Calculation of the Kondo temperature}

As the Kondo screening process defines a crossover region in the phase diagram, the criterion 
to obtain the associated Kondo temperature must be precisely defined.
In the following we describe the specific algorithms used to extract the Kondo temperature $T_{\rm K}$ for the AIM and the PAM.
First, we illustrate the most ``conventional" one, based on the determination of the temperature dependence of the static local 
magnetic response. Second, we provide a thorough description on how to directly extract the value of $T_{\rm K}$ 
from the lowest frequency data of the generalized charge susceptibility.  

\begin{figure*}[tb]
\centering
{{\resizebox{8.8cm}{!}{\includegraphics {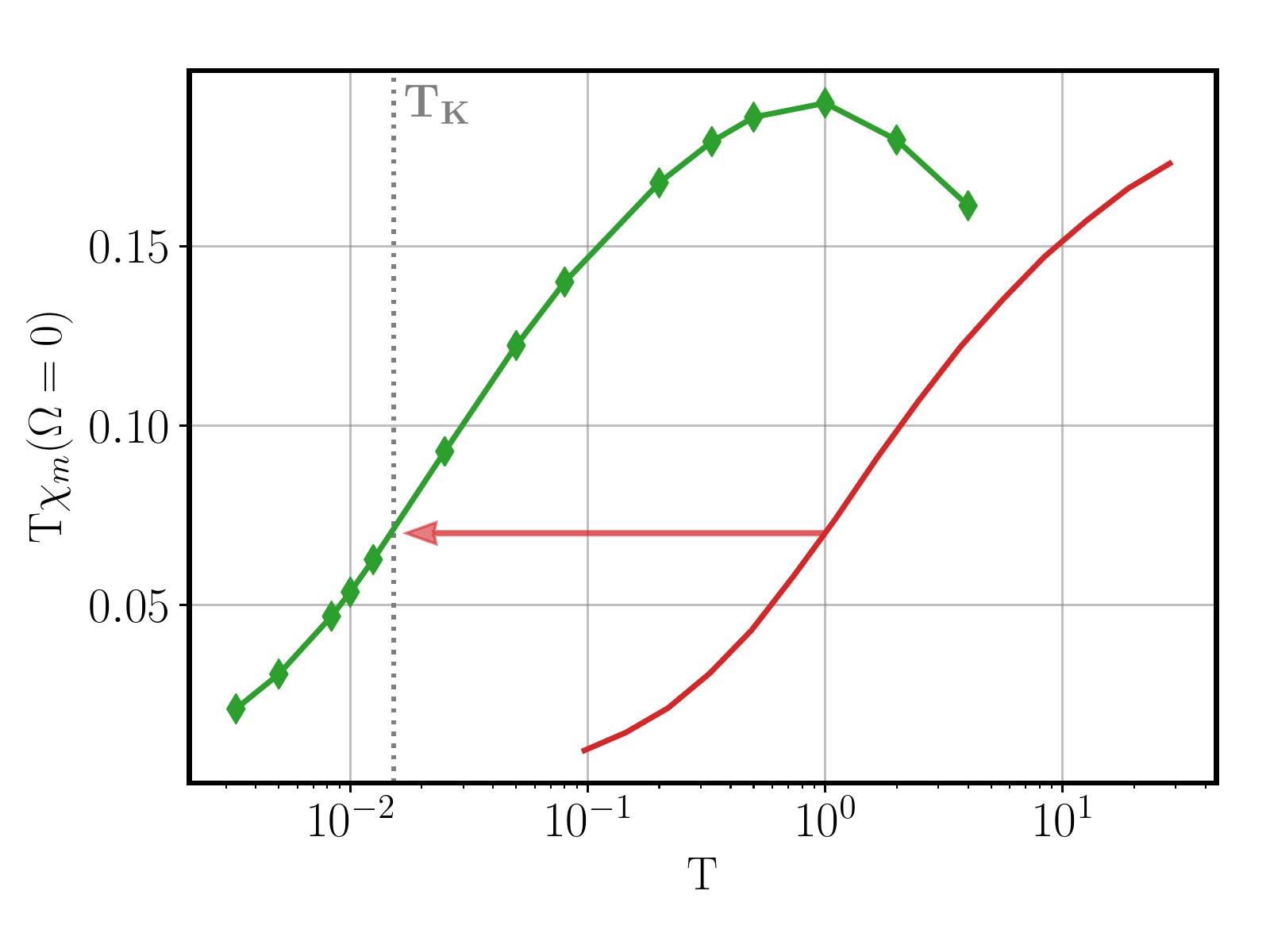}}}
{{\resizebox{8.8cm}{!}{\includegraphics {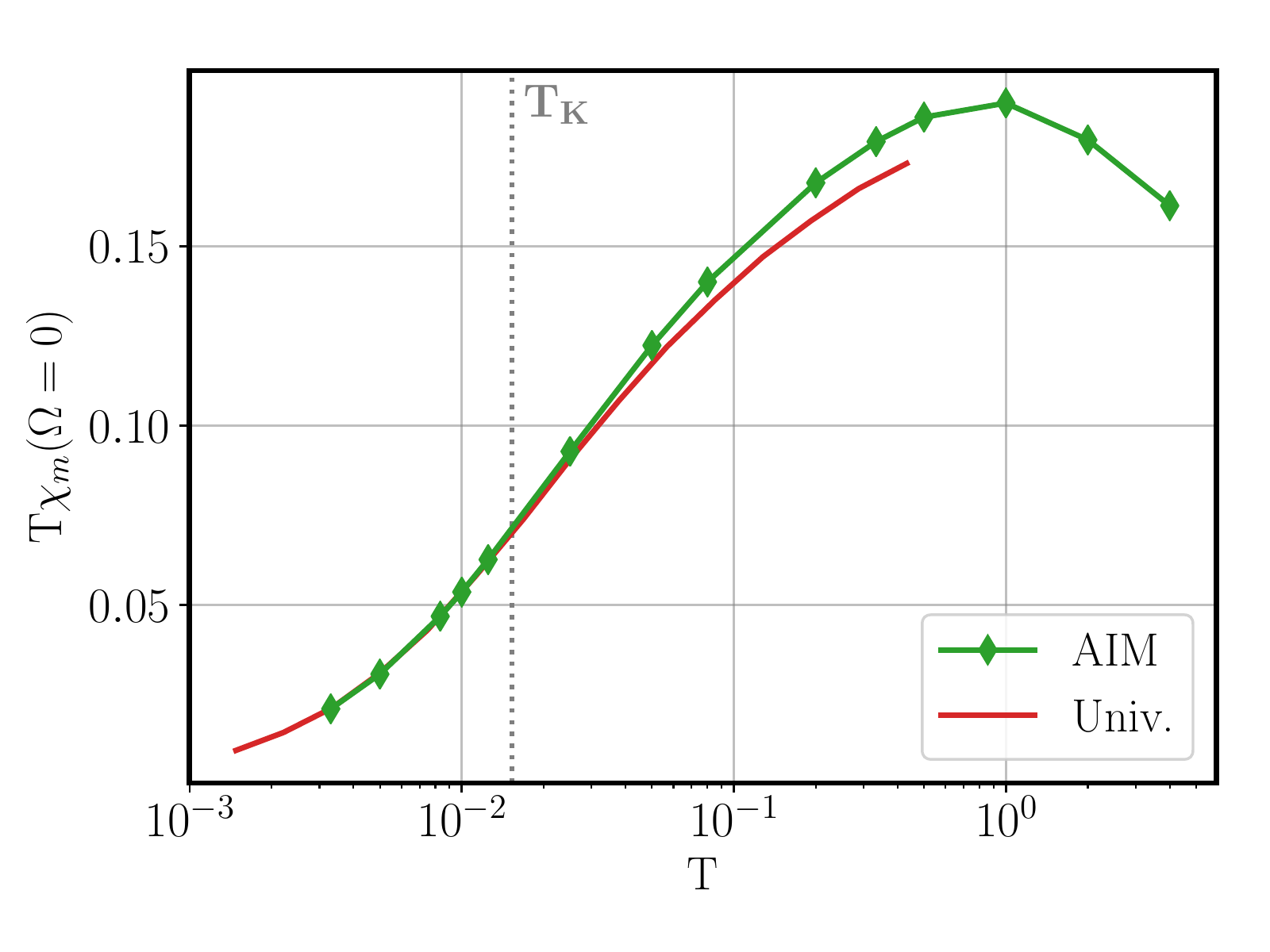}}}}
\caption{{\small
Extracting the value of $T_{\rm K}$ of the AIM for a given interaction value ($U=5.75$) from the numerical 
data for the static local magnetic susceptibility 
$\chi_{m}(\Omega=0)$ (QMC, green), by shifting the universal result (red) onto it (highlighted by the red arrow), obtained 
from the renormalization-group solution of a Kondo-Hamiltonian\cite{Krishnamurthy1975,Krishnamurthy1980}. The value for $T_{\rm K}$ obtained 
in this way is shown as the vertical grey dotted line.}}
\label{fig:TK-1}  
}
\end{figure*}

\subsection{T-dependence of the magnetic susceptibility}
\label{Tk-Sec}

We have determined the precise value of the Kondo temperature $T_{\rm K}$ for the AIM from the overall 
temperature dependence of the static magnetic susceptibility $\chi_{m}(\Omega=0)$ on the impurity site. 
This (well-known) procedure was described in a work by H. R. Krishna-Murthy~[\onlinecite{Krishnamurthy1975,Krishnamurthy1980}] and is also summarized in 
Ref.~[\onlinecite{Hewson1993}]. It works by comparing the temperature evolution of $\chi_{m}(\Omega=0)$ for a fixed 
interaction value $U$ 
to a universal renormalization group solution for a Kondo-Hamitonian\cite{Krishnamurthy1975,Krishnamurthy1980}\footnote{Note that the constants used in Refs.~\cite{Krishnamurthy1975,Krishnamurthy1980}
are included in our definition of the static magnetic susceptibility, leading to $\chi_m=\nicefrac{\chi^{\text{Refs.~\onlinecite{Krishnamurthy1975,Krishnamurthy1980}}}_m}{(g\mu_B)^2}$, where $g=2$ and $\mu_{\rm B}$ is the Bohr magneton.}.
In practice, one must ($i$) compute $T\chi_{m}(\Omega=0)$ in a quite large temperature range and 
($ii$) shift the data of the universal result\cite{Krishnamurthy1975,Krishnamurthy1980}, plotted as a function of $\log({T/{T_{\rm K}}})$ with $T_{\rm K}=1$, onto the 
numerical result for $T\chi_{m}(\Omega=0)$. This way one obtains the Kondo temperature for the AIM for this value of the interaction $U$. 
The procedure 
is shown in Fig.~\ref{fig:TK-1} for $U=5.75$, where the unshifted case is plotted in the left panel, the shifted one in the right. 
This shift is applied in such a way that the agreement between the universal result for the Kondo-Hamiltonian (red) and the numerical one for the AIM (green) 
is the most precise for low temperatures $T\lesssim T_{\rm K}$.
Using this procedure, the value of $T_{\rm K}$ for the specific AIM used throughout this work was obtained for several interaction values, shown as 
blue crosses in the main text in Fig.~2. The same method was already used in an earlier work, see Appendix B of Ref.~[\onlinecite{Chalupa2018}]. In this work the resulting $T_{\rm K}$ 
was also compared to an analytic result\cite{Hewson1993} 
obtained for the wide-band limit of the AIM ($D\gg U$), yielding an excellent agreement.

For the PAM the maximum of $\chi_m(\Omega=0)$ as a function of the temperature was used to determine the value of the Kondo 
temperature, see further the supplemental material of Ref.~[\onlinecite{Schaefer2019}].

\subsection{Low-frequency criterion from ${\tilde{\chi}^{\nu\nu'}}$}

\begin{figure}[tb]
\centering
{{\resizebox{8.8cm}{!}{\includegraphics {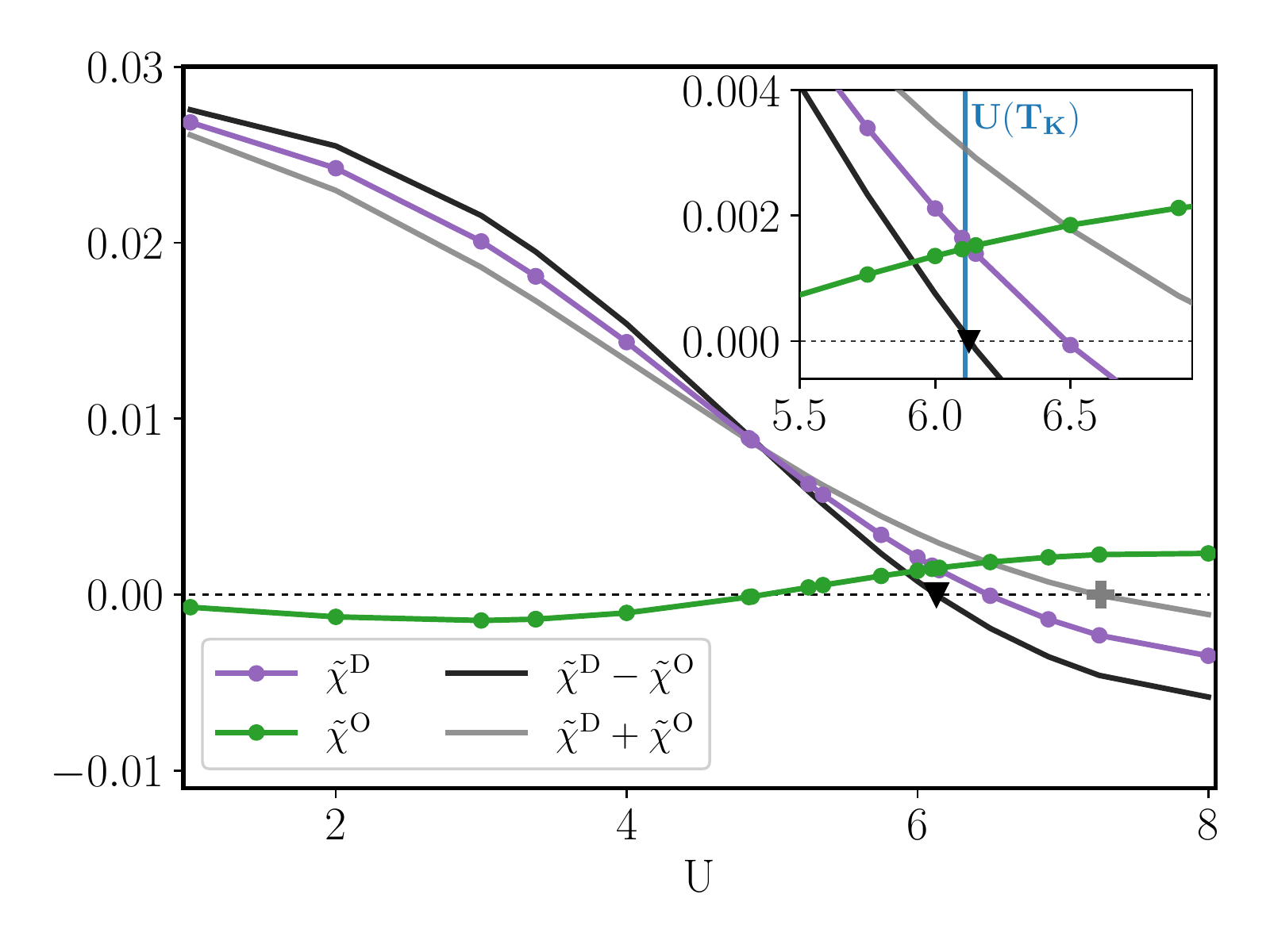}}}
\caption{{\small 
Behavior of $\tilde{\chi}^{\rm D}$ (violet) and $\tilde{\chi}^{\rm O}$ (green) as a function of $U$ for $T=0.0125$ for the AIM. 
The black line represents the first eigenvalue of the $2\times 2$ submatrix of $\tilde{\chi}^{\nu \nu'}$ ($\tilde{\chi}^{\rm D}-\tilde{\chi}^{\rm O}$), which 
is singular (black triangle) closely to ${\rm U(T_{\rm K})}$. 
The second low-frequency criterion obtained from the second eigenvalue ($\tilde{\chi}^{\rm D}+\tilde{\chi}^{\rm O}$) is 
shown as a grey line. 
}}
\label{fig:TK-2}  
}
\end{figure} 

As we discussed in the main text, at $T\approx T_{\rm K}$ the generalized charge susceptibility acquires 
a typical ``onion"-structure. Beyond this qualitative feature, the value of $T_{\rm K}$ can be extracted for 
large interaction values, by a precise condition on the lowest frequency entries of $\tilde{\chi}^{\nu\nu'}$: 
$\tilde{\chi}^{\rm D} = \tilde{\chi}^{\rm O}$, where $\tilde{\chi}^{\rm D} = T^2\tilde{\chi}^{\,\pi T, \pi T}$ 
and $\tilde{\chi}^{\rm O} = T^2\tilde{\chi}^{\,\pi T, -\pi T}$.
We note in passing, that a practical quality of this 
criterion resides in the 
possibility of performing a bisection. 

Since the singularity of the innermost $2\times 2$ submatrix of $\tilde{\chi}^{\nu \nu'}$ 
is a precise reference-point on the two-particle level, $T_{\rm K}$ can be obtained either by a scan in temperature 
for fixed interaction values (as shown in the lower panel of Fig.~1 in the main text), or vice versa. The second 
possibility is shown 
in Fig.~\ref{fig:TK-2} of this supplemental material, where the temperature is fixed to $T=0.0125$, and the interaction 
value is varied in a broad range. By monitoring the value of $\tilde{\chi}^{\rm D}-\tilde{\chi}^{\rm O}$ (black line) 
as a function of $U$ one can readily identify the 
singularity of the $2\times 2$ submatrix of $\tilde{\chi}^{\nu\nu'}$ 
(black triangle), see the inset, where a zoom around $\rm{U(T_{\rm K})}$ (blue vertical line) is shown. The value $\rm{U(T_{\rm K})}$ refers in this context to the interaction 
value where $T_{\rm K}$ is equal to the temperature $T=0.0125$.

\begin{figure}[tb]
\centering
{{\resizebox{8.8cm}{!}{\includegraphics {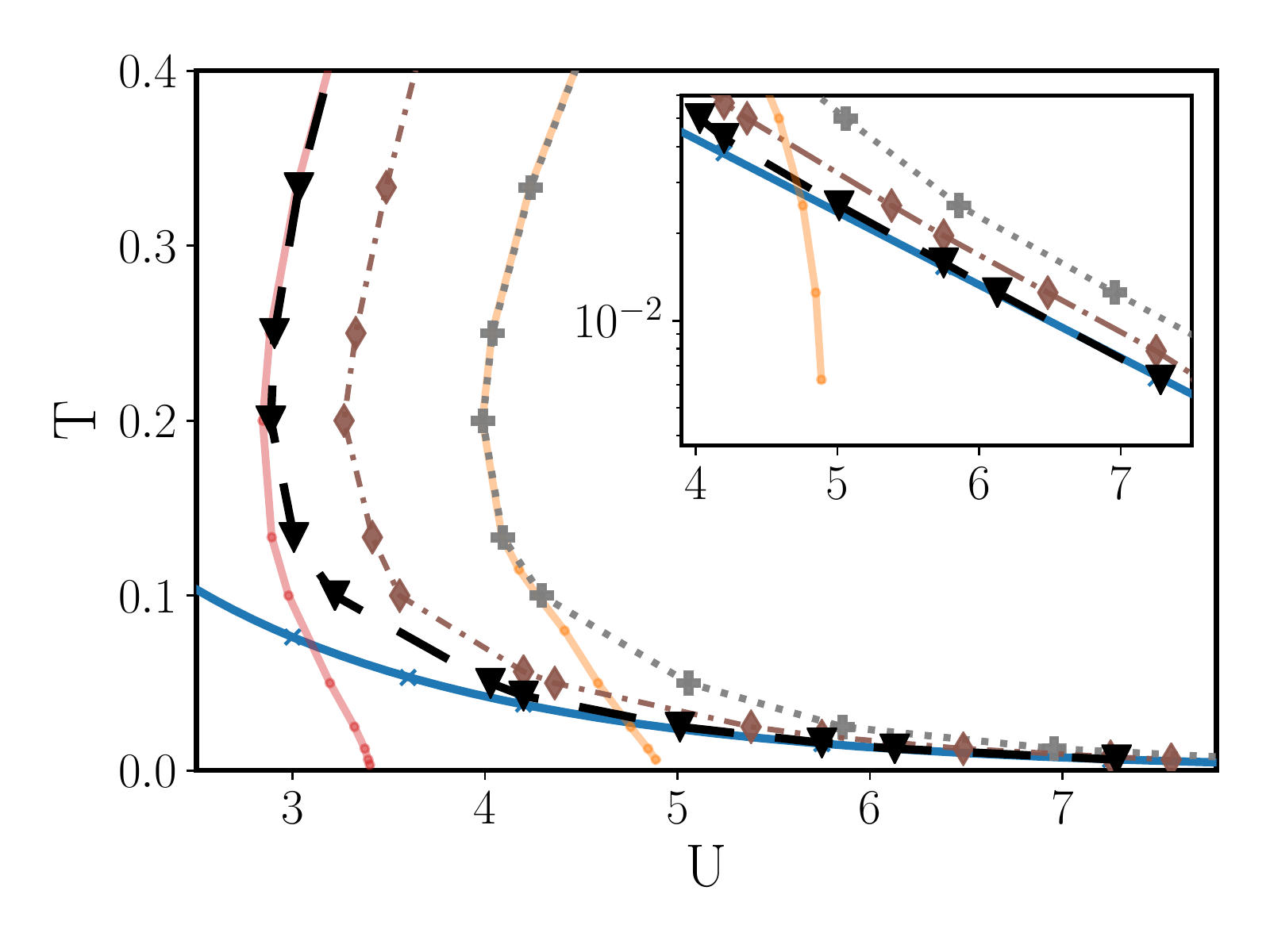}}} 
\caption{{\small 
$T-U$ diagram of the AIM, the blue solid line represents the Kondo temperature $T_{\rm K}$. 
Several low-frequency criteria are shown: ($i$) $\tilde{\chi}^{\rm D} \!=\! \tilde{\chi}^{\rm O}$ (black triangles), 
($ii$) $\tilde{\chi}^{\rm D} \!=\! -\tilde{\chi}^{\rm O}$ (grey plus symbols) and ($iii$) $\tilde{\chi}^{\rm D} \!=\! 0$ (brown diamonds). 
Only the first criterion ($i$) lies on-top of $T_{\rm K}$ 
for large interaction values and low-temperatures, as discussed in the main text. The others are close, but not 
on-top, see the logarithmic inset. The first and the second vertex 
divergence line for the AIM are shown as red and orange line, respectively, which coincide
with conditions ($i$) and ($ii$) for high-temperatures. }}
\label{fig:TK-3}  
}
\end{figure} 

As mentioned in the main text, the condition $\tilde{\chi}^{\rm D}=\tilde{\chi}^{\rm O}$ turns out to be the most accurate one to match the value 
of $T_{\rm K}$ for large interaction values and low-temperatures. 
Here we compare the criterion ($i$) $\tilde{\chi}^{\rm D}=\tilde{\chi}^{\rm O}$ to other 
reasonable low-frequency criteria one could think of, in particular 
($ii$) $\tilde{\chi}^{\rm D}=-\tilde{\chi}^{\rm O}$ and ($iii$) $\tilde{\chi}^{\rm D}=0$. The results of this comparison are 
shown in Fig.~\ref{fig:TK-3}. Here the second singularity of the $2\times 2$ submatrix, $\tilde{\chi}^{\rm D}=-\tilde{\chi}^{\rm O}$, is represented 
by the grey plus symbols. 
We note that, for high temperatures, this coincides with the singularity of the whole matrix
$\tilde{\chi}^{\nu \nu'}$, which leads to the second vertex divergence observed in the AIM\cite{Chalupa2018}, 
shown in orange (color code 
compatible with earlier works on vertex divergences\cite{Schaefer2013,Schaefer2016c,Chalupa2018,Thunstroem2018,Springer2019}). 
At the same time, the brown diamonds denote the parameter set where $\tilde{\chi}^{\rm D}=0$ holds, which 
lies in between ($i$) and ($ii$). As one notices readily in the inset shown in Fig.~\ref{fig:TK-3} all these low-frequency
criteria are fairly close 
to the Kondo temperature for large interaction values and low-temperatures. However they can be clearly distinguished
from $\tilde{\chi}^{\rm D}=\tilde{\chi}^{\rm O}$, which lies {\sl on-top} of $T_{\rm K}$.

\section{Frequency structures in ${\tilde{\chi}^{\nu\nu'}}$}

In the main text the characteristic frequency structures of the local moment formation as well as of its Kondo screening, the so-called {\sl fingerprints}, are identified for the AIM. 
In this part of the supplemental material, we provide further details on the underlying nonperturbative frequency structures, and we also present the corresponding results for the PAM and the HM.

\subsection{Relation to the physical charge response $\chi$}

The connection between the frequency structure of $\tilde{\chi}^{\nu\nu'}$ and the corresponding 
behavior of the physical response $\chi$  in the local moment and the Kondo regime, as discussed in the main text, can be also traced in the results of partial summations of the generalized charge susceptibility.
Specifically, we consider the $\nu,\nu'$ summation of $\tilde{\chi}^{\nu\nu'}$ over frequency boxes of increasing sizes, as detailed by the following expression:

\begin{equation}
    \chi_{\rm partial} (\nu_{\rm max}) = T^2 \sum\limits_{\nu,\nu'= -\nu_{\rm max}}^{\nu_{\rm max}} \, \tilde{\chi}^{\, \nu\nu'}
    \label{eq:chipartial}
\end{equation}

Evidently, for $\nu_{\rm max} \rightarrow \infty$, $ \chi_{\rm partial}$ corresponds to $\chi$, and Eq.~(\ref{eq:chipartial}) of this supplemental material reduces to Eq. (2) of the main text. 
At the same time, by inspecting the results obtained for finite $\nu_{\rm max}$, the energy-selective effects of different nonperturbative/perturbative features in $\tilde{\chi}^{\nu\nu'}$  can be clearly individuated.

\begin{figure}[t]
\centering
{{\resizebox{8.8cm}{!}{\includegraphics {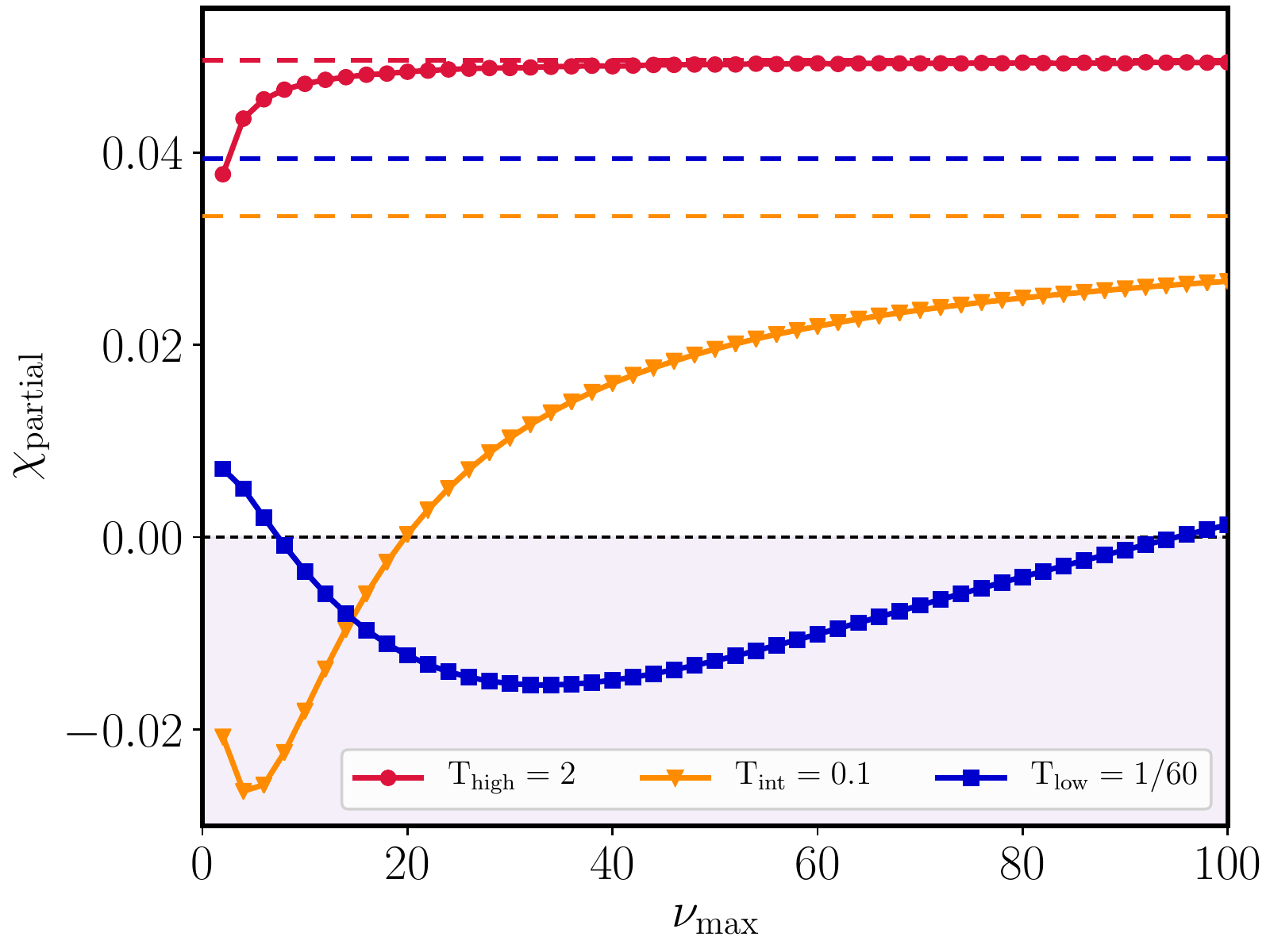}}} 
\caption{{\small Partial frequency summation of $T^2\sum_{\nu \nu^{\prime}} \tilde{\chi}^{\nu \nu^{\prime} }$, see Eq.~(\ref{eq:chipartial}), over 
frequency boxes of increasing size ($\nu_{\rm max}$) for three different temperatures regimes (solid lines). 
The values of the physical response $\chi$, which correspond to the $\nu_{\rm max}\!\rightarrow\!\infty$-limit, are depicted as dashed lines. 
The area where $\chi_{\rm partial }$ is negative is highlighted by a grey background.
 }}
\label{fig:Nonpert-1}  
}
\end{figure}

In Fig.~\ref{fig:Nonpert-1} the result of this partial sum is displayed as a function of $\nu_{\rm max}$ for the three different temperatures regimes
illustrated in Fig.~1 of the main text ($T_{\rm high},T_{\rm int} $ and $ T_{\rm low}$). 

At high-temperature ($T_{\rm high}$), i.e. in the perturbative regime, $\chi_{\rm partial}$ (red circles) is always {\sl positive} and {\sl increases} monotonically toward its asymptotic value for $\nu_{max} \rightarrow \infty$ (red dashed line). 

For intermediate temperatures ($T_{\rm int}$) we enter the local moment regime of the AIM. As we discussed in the main text, this results in large negative diagonal entries in $\tilde{\chi}^{\, \nu\nu'}$ at low-to-intermediate frequencies. Due to these negative contributions, $\chi_{\rm partial}$ (orange triangles) is {\sl negative} for small $\nu_{\rm max}$ and further {\sl decreases} until, for a certain value of $\nu_{\max}$ a minimum is reached. 
Thereafter, the perturbative high-frequency
asymptotics comes into play, slowly enhancing $\chi_{\rm partial}$  until a positive value of $\chi$ is eventually 
obtained for $\nu_{\rm max} \rightarrow \infty$ (orange dashed line). Because of the initial {\sl negative} low-frequency contributions to $\chi_{\rm partial}$, the final value of the physical charge response $\chi$ 
gets strongly suppressed w.r.t.~the perturbative one.

At low-temperatures ($T_{\rm low}$) in the Kondo regime $\tilde{\chi}^{\nu \nu^{\prime} }$ displays the characteristic ``onion-structure" illustrated in the main text. This represents a clear hallmark of the competition between the two trends discussed above. $\chi_{\rm partial}$ (blue diamonds) starts {\sl positive}, due the positive low-frequency elements of the diagonal of $\tilde{\chi}^{\nu \nu^{\prime}}$, which can be seen in the right central panel of Fig.~1 within the white square. $\chi_{\rm partial}$ then {\sl decreases} as a function of $\nu_{\rm max}$ until a minimum is reached, similarly as in the local moment regime. Eventually it increases again until a positive value is recovered for $\nu_{\rm max} \rightarrow \infty$. Due to the initial positive sign of the low-energy contributions to the frequency sum of Eq.~(\ref{eq:chipartial}) of this supplemental material, the final value of the physical response $\chi$ gets slightly {\sl enhanced} w.r.t. the corresponding one in the local moment regime, where these positive low-energy contributions to the partial sum are absent.  

\subsection{Correlation of particles with antiparallel spins}

\begin{figure}[t]
\centering
{{\resizebox{8.8cm}{!}{\includegraphics {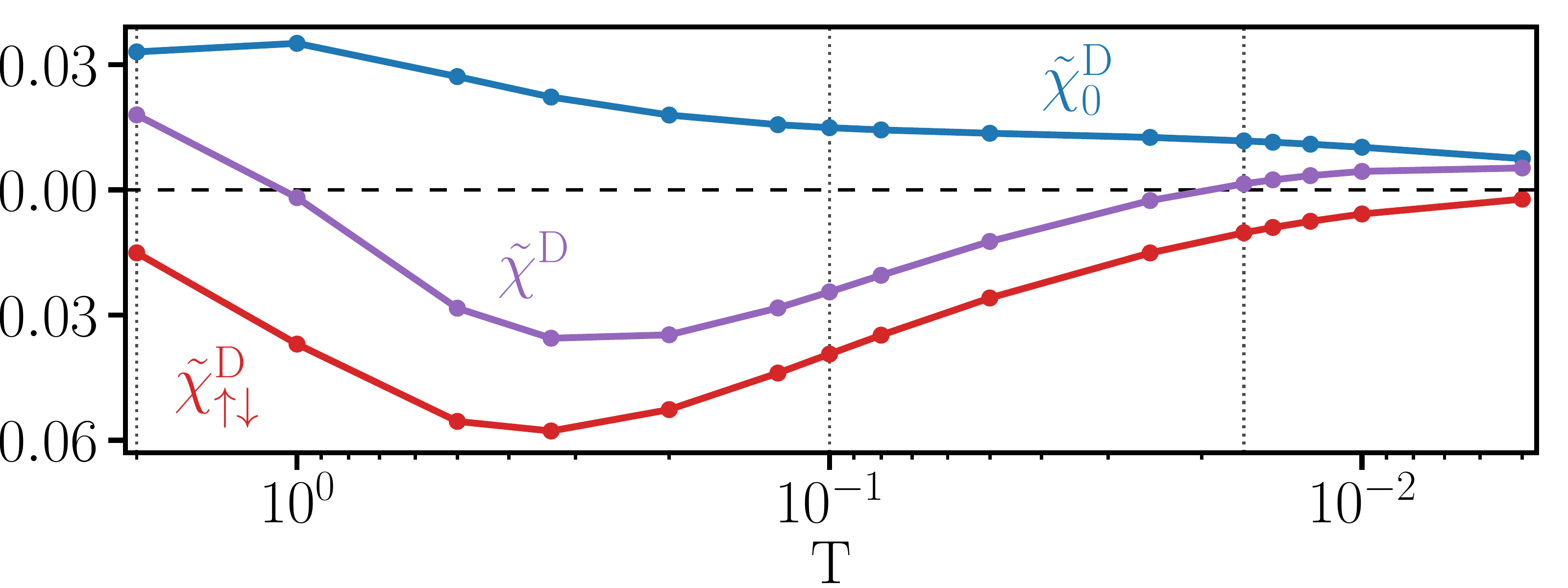}}} 
\caption{{\small Behavior of the lowest-frequency diagonal element of $T^2\tilde{\chi}^{\nu=\nu'}$ for the $\uparrow\uparrow$ (blue), the $\uparrow\downarrow$ (red) and the charge ($\uparrow\uparrow + \uparrow\downarrow$) 
sector (violet, also shown in Fig.~1 of the main text), as a function of the temperature on a logarithmic axis. The vertical dotted lines mark the three temperatures also shown in Fig.~1. }}
\label{fig:Nonpert-2}  
}
\end{figure} 

Here we show the data of the lowest-frequency diagonal element of the generalized susceptibility for particles with antiparallel spins, 
$T^2\tilde{\chi}^{\, \pi T, \pi T}_{\uparrow \downarrow} = \tilde{\chi}^{\rm D}_{\uparrow \downarrow}  $,
as a function of the temperature $T$, for the same parameter set used in Fig.~1 in main text.
In our related discussion of the onion-structure, we mention that the weakening of correlations in this sector is the main cause for the behavior of $\tilde{\chi}^{\rm D}$.
This can be readily verified in Fig.~\ref{fig:Nonpert-2}, where as a comparison also $\tilde{\chi}^{\rm D}_0$ (blue line) is shown, which is the only remaining contribution to $\tilde{\chi}^{\rm D}$ (violet). Note 
that this is due to the fact, that there are no vertex corrections on the diagonal in the $\uparrow\uparrow$ sector\cite{Rohringer2012,Rohringer2013a}, leaving only the bubble term for $\nu=\nu'$: 
$ T^2\tilde{\chi}_{\uparrow \uparrow}^{\nu = \nu'}  \!= \! T^2\tilde{\chi}_0^{\nu \nu'} \!=\! - T\delta_{\nu \nu'} G(\nu)^2$.

\subsection{Results for the PAM and the HM}

As discussed in the main text, the local moment formation as well as the Kondo screening are reflected in the frequency structure of $\tilde{\chi}^{\nu \nu'}$.

\begin{figure}[tb]
\centering
{{\resizebox{8.98cm}{!}{\includegraphics {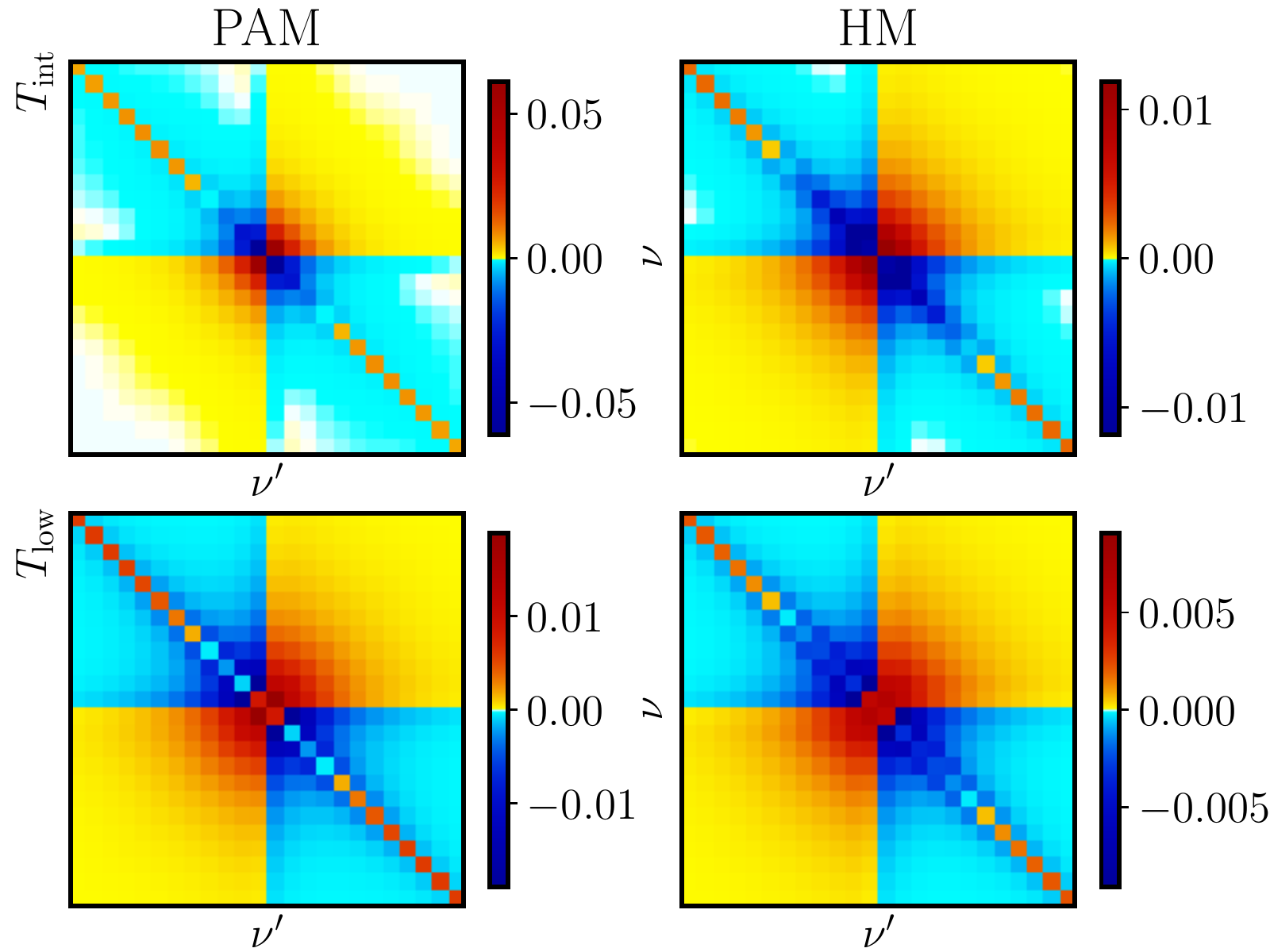}}} 
\caption{{\small 
Comparison of the frequency structure of $T^2\tilde{\chi}^{\nu\nu'}$ for the PAM (left column) and the HM (right column). At intermediate temperatures $T_{\rm int}>T_{\rm K}$ (top panels) one observes the local moment structure, described in the main 
text for the AIM (see Fig.~1 therein). The same holds for the
$T_{\rm low}\approx T_{\rm K}$ regime (bottom panels), where one recognizes for both cases the characteristic ``onion-structure". }}
\label{fig:TK-4}  
}
\end{figure}

The specific features, the so-called {\sl fingerprints},
can also be observed for the PAM and the HM, as shown in Fig.~\ref{fig:TK-4}. The top panels display the results for
the generalized charge susceptibility $\tilde{\chi}^{\nu\nu'}$ (normalized by $T^2$) at intermediate temperatures ($U/D/V > T_{\rm int} >T_{\rm K}$), the 
bottom panels nicely show 
the ``onion-structure" observed for both, the PAM (left column) and the HM (right column), in the Kondo regime 
($T_{\rm low}\approx T_{\rm K}$).

\section{Comparison to perturbative methods}


As discussed in the main text, perturbative methods fail in accounting for the impact of
the physics of the magnetic channel 
(i.e. the local moment formation and its screening) onto the charge channel, see Fig.~3 of the
main text. This roots back to their lack of describing vertex divergences. 

In this supplemental material we substantiate the statements made in the main text with more details. 
We briefly describe the PA and fRG method used for our comparison,  
and then discuss to which extent the qualitative behavior of the magnetic channel is correctly reproduced by these two methods.

\begin{figure*}[tb]
\centering
{{\resizebox{8.8cm}{!}{\includegraphics {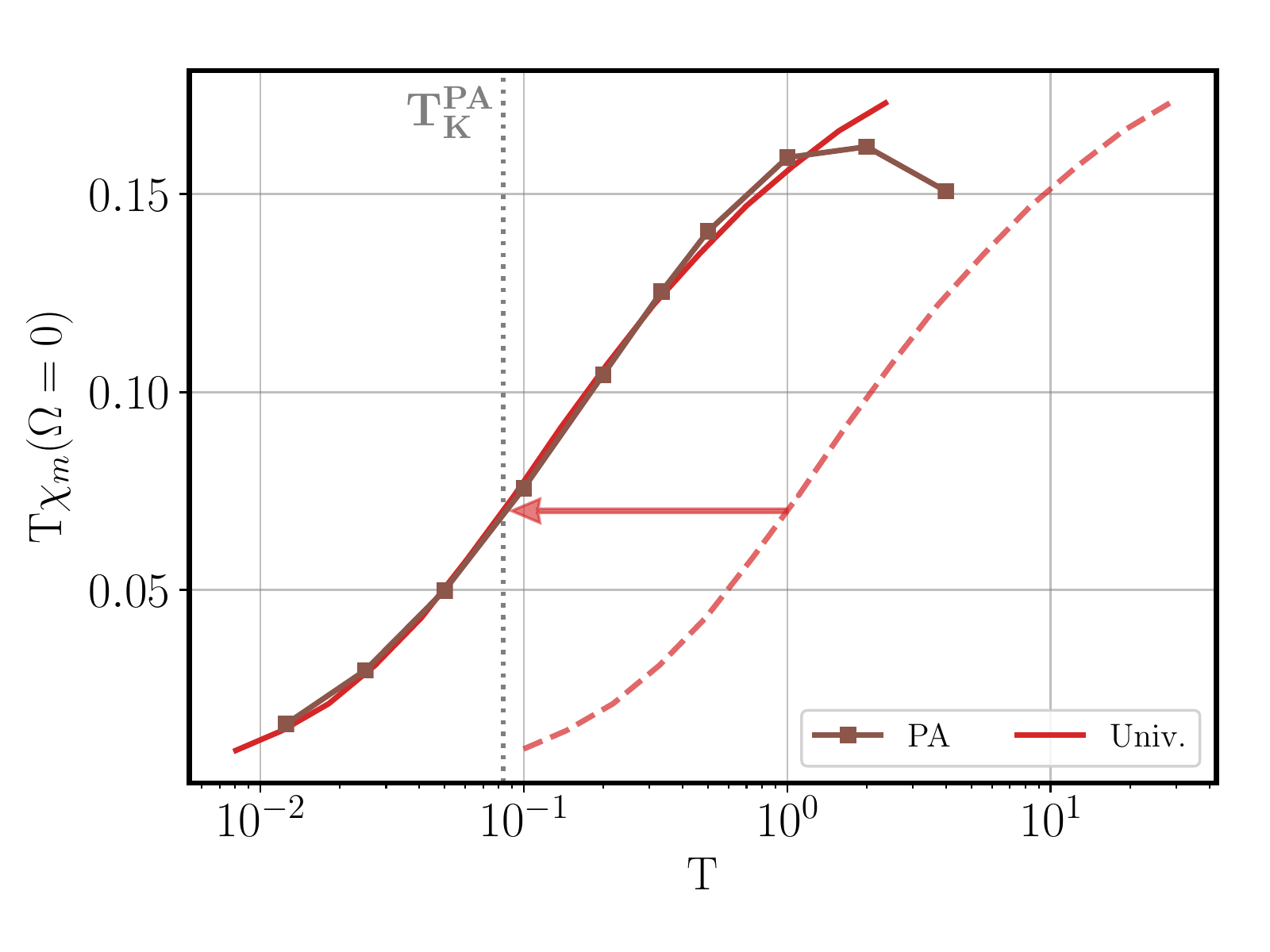}}}
{{\resizebox{8.8cm}{!}{\includegraphics {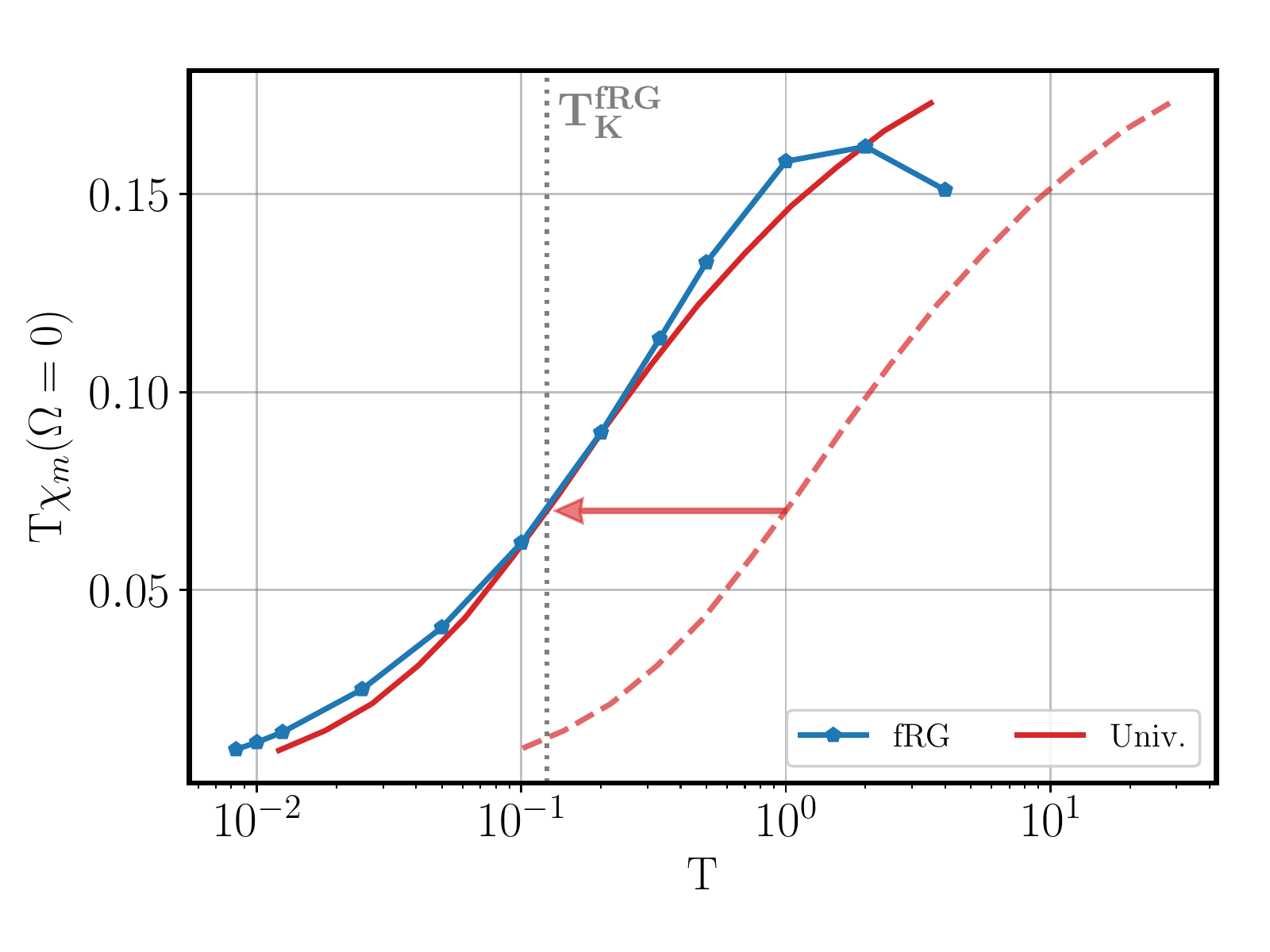}}} }
\caption{{\small 
Extracting $T_{\rm K}$ in the standard way described in Sec.~\ref{Tk-Sec}, for the PA (left) and the fRG (right) data for $U=4.2$, where the respective result is shown as vertical dotted line for each method. 
Interestingly, the PA manages to describe the
qualitatively correct behavior of $\chi_{m}(\Omega=0)$, albeit with a too high $T^{\rm \, PA}_{\rm K}\gtrsim 2 T_{\rm K}$.
The fRG method with a 1$\ell$-truncation, gives an even larger estimate for $T_{\rm K}$ and displays 
some qualitative deviations for $T<T_{\rm K}$.  }}
\label{fig:PAfRG-1}  
}
\end{figure*} 

\subsection{Details of the fRG method}
\label{fRG-Details}

The starting point of the fRG\cite{Metzner2012} is an exact functional flow equation, which yields the gradual evolution from a microscopic model action to the final effective action as a function of a flowing scheme-dependent regulator. By expanding in powers of the fields one obtains an exact hierarchy of flow equations for vertex functions, which is in practical implementations restricted to the one- and two-particle vertex. 
The underlying approximations, often at the one-loop (1$\ell$) level, are devised for the weak to moderate coupling regime\footnote{More strongly correlated parameter regimes might become accessible by exploiting the DMFT as a starting point for the (multiloop) fRG flow\cite{Taranto2014,Vilardi2019}.}, where forefront algorithmic advancements brought the fRG for interacting fermions on 2D lattices to a quantitatively reliable level\cite{Tagliavini2019,Hille2020}. 

The fRG approach has been applied earlier to study Kondo physics in and out-of equilibrium within the single-impurity Anderson model and more complex variants of the latter\cite{list1A,list1B,list1C,list1D,list1E,list1F,list1G,list1H,list1I,list1J,list1K,list1L,list1M,list1N,list1O,list1P}. The static truncation of the first implementations was subsequently extended to account for all second order processes including a frequency dependent two-particle vertex and self-energy, taking into account the full real-space as well as spin structure\cite{list1F,list1H,list1L,list2A}. Including the frequency dependence clearly improves the results beyond bare perturbation theory but it does not allow to reach the strong coupling regime in a controlled way\cite{Metzner2012}.

The numerical results presented in this work have been obtained by employing a $1\ell$-scheme using the Katanin replacement\cite{Katanin2004} in the flow equation for the two-particle vertex.
The accuracy of our treatment is higher with respect to previous fRG-based implementations since the complete
frequency dependence is taken into account. In particular, we use a full frequency treatment of all three independent dependencies at low frequencies together with a refined scheme for the high-frequency asymptotics\cite{Rohringer2012,Wentzell2016}.

We refer to \cite{Wentzell2016,Tagliavini2019,Chalupa2020} for the details of the algorithmic implementation.


\subsection{Details of the PA method}
\label{PA-Details}

The parquet approximation (PA) is based on the parquet equations\cite{DeDominicis1964, DeDominicis1964b, Janis1999, Bickersbook2004}, 
which is a system of exact equations relating reducible and irreducible 
two-particle vertices. 
The irreducibility on the two-particle level is defined with respect to cutting two fermionic lines\cite{Abrikosov1975,Bickersbook2004,Rohringer2012}. 
Together with the Dyson equation and the Schwinger-Dyson equation, relating the full vertex with the one-particle 
self-energy, one can obtain all quantities of interest staring from a given input for the 
fully irreducible vertex. 
In the PA this fully irreducible vertex is approximated by its lowest order term, i.e. the bare interaction U\cite{Bickersbook2004,Yang2009,Valli2015,Li2016}.
Let us note that we restrict ourselves to the conventional version of the PA, {\sl i.e.} not including the recent modifications
illustrated in Refs.~\cite{Janis2007,Janis2008,Janis2019,Janis2019JPS,Janis2020}.

In our calculations, we used an iterative method to solve the parquet equations, as also described in 
Ref.~[\onlinecite{Wentzell2016}]. This is implemented in the same framework as the 
fRG method described above guaranteeing a certain degree of comparability between the
numerical calculations of the two approaches.

\begin{figure}[tb]
\centering
{{\resizebox{8.8cm}{!}{\includegraphics {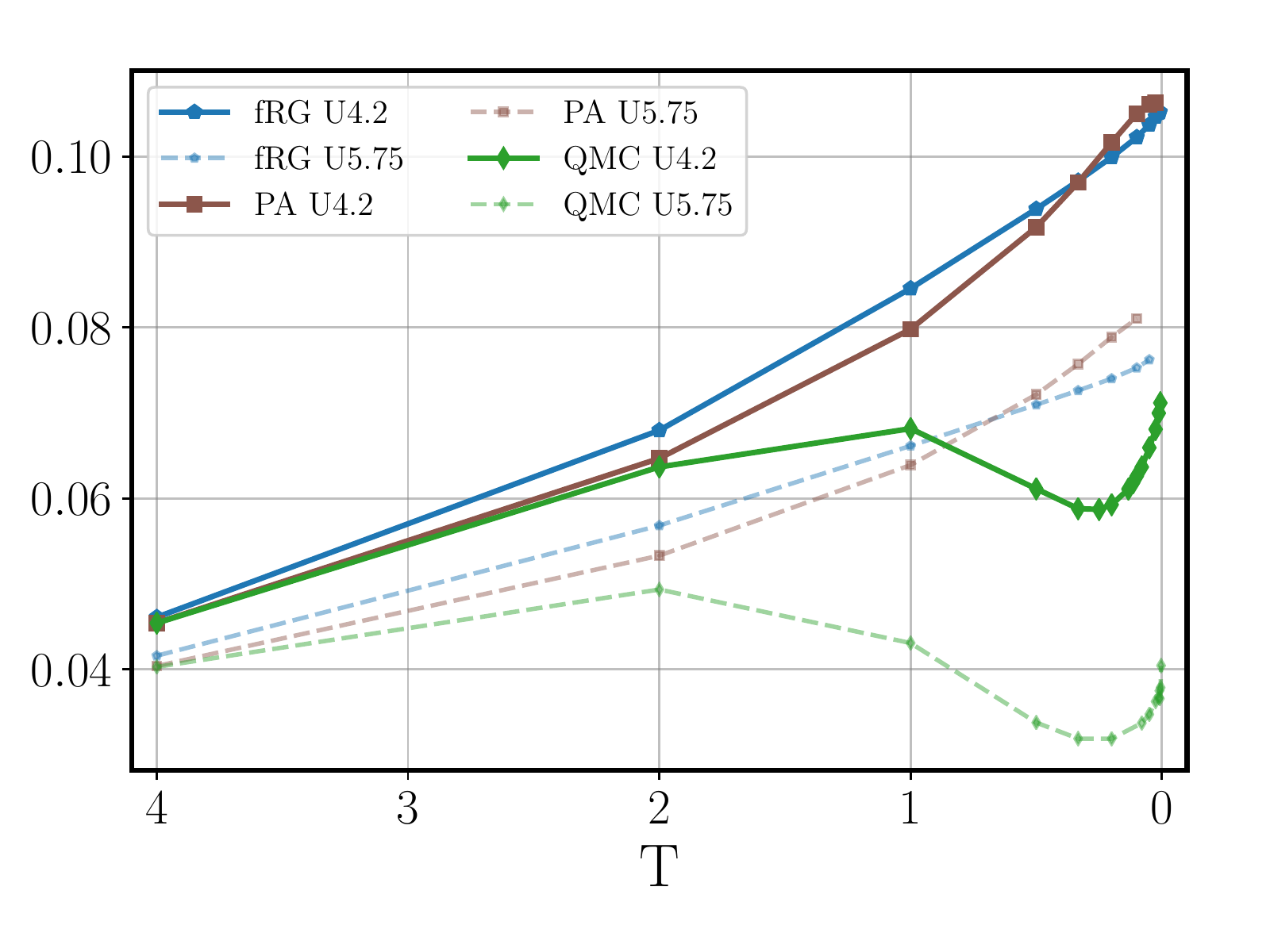}}}
\caption{{\small 
Temperature behavior of the physical static charge susceptibility $\chi$ for $U=4.2$ (solid) and $U=5.75$(dashed) for 
the exact solution (QMC, green diamonds) and perturbative approaches (PA, brown squares and fRG, blue pentagons). }}
\label{fig:PAfRG-2}  
}
\end{figure}

\subsection{Static magnetic susceptibility in fRG and PA}

Here we discuss fRG and PA results for $\chi_{m}(\Omega=0)$. 
We perform a temperature scan for $U=4.2$ and extract the corresponding Kondo temperature $T_{\rm K}$, 
in the same manner as discussed in Sec.~\ref{Tk-Sec} of this supplemental material. The interaction value $U=4.2$ 
was chosen such that the low-temperature regime ($T < T_{\rm K}$) can be still accessed numerically with our PA solver, 
differently from the $U=5.75$ case shown in the main text.

In the left panel of Fig.~\ref{fig:PAfRG-1} the Kondo temperature is extracted from the PA data (brown). 
The corresponding result  
is incorrect at the quantitative level ($T^{\rm \, PA}_{\rm K} = 1/12 \gtrsim 2T_{\rm K}$, where $T_{\rm K}=1/27$)\footnote{Note that a similar trend for the estimate 
of the Kondo temperature has been reported in the reduced two-particle self-consistent parquet approach of Ref.~\cite{Janis2019,Janis2019JPS,Janis2020}}.
Nonetheless, the PA method yields a qualitatively correct 
description of the local moment physics (namely of its overall temperature-dependence) in the magnetic
channel. 
Similarly as in Fig.~\ref{fig:TK-1} the unshifted universal result is shown as the dashed red line, whereas the arrow illustrates the shift. 
In the same way the fRG case (blue) is shown in the right panel of Fig.~\ref{fig:PAfRG-1}.
Here one notices that, as in the PA case, the quantitative value is  
too large ($T^{\rm \, fRG}_{\rm K} = 1/8 $). Further the 
qualitative description is no longer perfect, showing deviations from the universal result for $T<T_{\rm K}$.
Let us note at this point, that a recent extension of the 1$\ell$-fRG, the multiloop-fRG\cite{Kugler2018,Kugler2018b} (mfRG), 
contains all PA diagrams. This means that, if it is solved, the result is regulator-independent and coincides with the PA solution, removing the 
low-$T$ discrepancies observed in Fig.~\ref{fig:PAfRG-1}.

\begin{figure}[t]
\centering
{{\resizebox{8.8cm}{!}{\includegraphics {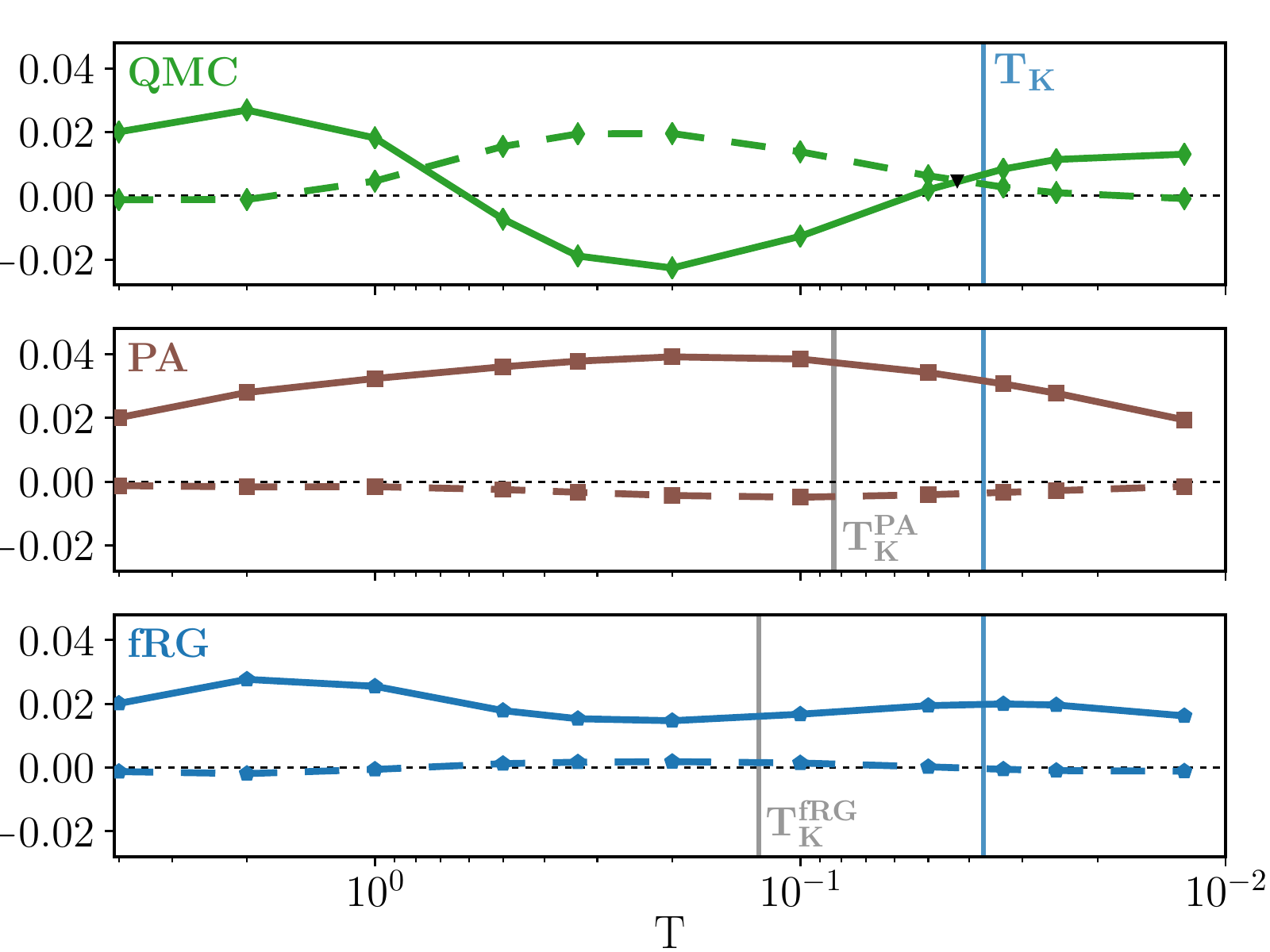}}} 
\caption{{\small Comparison of  $\tilde{\chi}^{\rm D}$ and $\tilde{\chi}^{\rm O}$ for the QMC (top), PA (middle), fRG (bottom) for $U=4.2$ and different temperatures. 
Also shown as a vertical blue line in all panels is the Kondo temperature $T_{\rm K}$ as obtained from the numerically exact solution for $U=4.2$.
The grey vertical lines show the Kondo temperature as obtained from the approximative solutions (PA and fRG) described above. 
 }}
\label{fig:PAfRG-4}  
}
\end{figure} 

Summarizing, both perturbative methods describe the physics of the magnetic channel in a qualitatively correct way 
(with somewhat larger 
deviations in the 1$\ell$-fRG case). 
This contrasts entirely their performance in the charge sector, where their description is fully incorrect, 
even at a qualitative level, as shown and discussed in the main text (see Fig.~3 therein).
The false description of the impact of the local moment formation and its screening 
onto the charge fluctuations roots back to the absence of divergences of the corresponding 
irreducible vertex functions\cite{Schaefer2013,Janis2014,Schaefer2016c,Ribic2016,Vucicevic2018, Chalupa2018,Thunstroem2018,Springer2019} 
in both perturbative approaches. The same consideration applies to the paring channel (not shown).
In Fig.~\ref{fig:PAfRG-2} the result for the physical local charge 
susceptibility $\chi$ is shown for $U=4.2$ (solid lines), as a reference also the 
result for $U=5.75$ of 
the main text is reproduced (dashed and transparent). In both cases it can readily noticed that {\sl both} perturbative methods fail in describing the 
minimum of $\chi(T)$ associated with the local moment regime of the AIM, as detailed in the main text.

To analyze this qualitative drawback in the description of local moment physics in greater detail, we provide a comparison for $U=4.2$ of the low-frequency elements of $\tilde{\chi}^{\nu\nu'}$, 
$\tilde{\chi}^{\rm D} = T^2\tilde{\chi}^{\, \pi T, \pi T}$ and $\tilde{\chi}^{\rm O} = T^2\tilde{\chi}^{\, \pi T, -\pi T}$, for all methods considered. 
Fig.~\ref{fig:PAfRG-4} shows the behavior of the diagonal frequency elements ($\tilde{\chi}^{\rm D}$) (solid lines) for the numerically exact QMC solution (green, top panel), 
the PA result (brown, middle) and the fRG result (blue, bottom), as well as of the corresponding off-diagonal ones ($\tilde{\chi}^{\rm O}$) (dashed lines).
The QMC data (top panel) for $U=4.2$ display qualitatively the same behavior as the QMC results for $U=5.75$ shown in Fig.~1 of the main text.
In particular one readily notices the negative sign of $\tilde{\chi}^{\rm D}$, which encodes the freezing of charge fluctuations in the local moment regime.
The corresponding data for the perturbative methods show instead qualitative differences w.r.t.~the QMC ones in the entire local moment regime as {\sl no} sign change in $\tilde{\chi}^{\rm D}$ (middle and bottom panels)
is observed in the whole temperature regime.
We also note that for the PA the largest deviations are found precisely for the temperatures where $\chi$ is mostly suppressed in the numerically exact solution. 
In conclusion, this comparison demonstrates unambiguously that it is the communication between the channels in the local moment regime, which is insufficiently described by the perturbative methods.

\end{document}